\documentclass[12pt,preprint,natbib209]{aastex}
\usepackage{amsfonts}
\begin{document}
\newcommand{\msol}{M$_{\odot}$}
\newcommand{\rsol}{R$_{\odot}$}
\newcommand{\freq}{$4.3^{+2.7}_{-1.2}\%$}
%
%
\title{The Frequency of Debris Disks at White Dwarfs}
\author{
Sara D. Barber\altaffilmark{1,5},
Adam J. Patterson\altaffilmark{1},
Mukremin Kilic\altaffilmark{1},
S. K. Leggett\altaffilmark{2},
P. Dufour\altaffilmark{3},
J. S. Bloom\altaffilmark{4},
D. L. Starr\altaffilmark{4}
}
\altaffiltext{1}{Homer L. Dodge Department of Physics and Astronomy, University of Oklahoma, 440 W. Brooks St., Norman, OK, 73019, USA}
\altaffiltext{2}{Gemini Observatory, 670 N. A'ohoku Place, Hilo, HI 96720, USA}
\altaffiltext{3}{D\'{e}partement de Physique, Universit\'{e} de Montr\'{e}al, C.P. 6128, Succursale Centre-Ville, Montr\'{e}al, Qu\'{e}bec H3C 3J7, Canada}
\altaffiltext{4}{Department of Astronomy, University of California, Berkeley, CA 94720, USA}
\altaffiltext{5}{barber@nhn.ou.edu}
%
%
\begin{abstract}
	We present near- and mid-infrared photometry and spectroscopy from PAIRITEL, IRTF, and \emph{Spitzer} of a metallicity-unbiased sample of 117 cool, hydrogen-atmosphere white dwarfs from the Palomar-Green survey and find five with excess radiation in the infrared, translating to a \freq~frequency of debris disks. This is slightly higher than, but consistent with the results of previous surveys. Using an initial-final mass relation, we apply this result to the progenitor stars of our sample and conclude that $1-7$M$_{\odot}$ stars have at least a 4.3\% chance of hosting planets; an indirect probe of the intermediate-mass regime eluding conventional exoplanetary detection methods. Alternatively, we interpret this result as a limit on accretion timescales as a fraction of white dwarf cooling ages; white dwarfs accrete debris from several generations of disks for $\sim$10Myr. The average total mass accreted by these stars ranges from that of 200km asteroids to Ceres-sized objects, indicating that white dwarfs accrete moons and dwarf planets as well as Solar System asteroid analogues.
\end{abstract}

%
%
\section{INTRODUCTION}
The study of extrasolar planetary systems has become a very active field that will continue to grow in the coming years and decades. Over the last two decades, several hundred exoplanets have been discovered, proving planet formation to be a robust process in the Galaxy. The vast majority of known exoplanets have been discovered using just two techniques; namely, radial velocity and transits. To date, radial velocity studies have revealed the presence of well over five hundred planets and the number of transiting systems has passed the two hundred mark \citep[see http://exoplanet.eu and][]{schneider}.

A cursory examination of these planet-hosting stellar systems reveals a curious trend. The frequency of Jovian planets increases from 3\%
for M dwarfs to 14\% for 2M$_{\odot}$ stars \citep{johnson,bowler}. In addition, the distribution of known exoplanets as a function of host stellar mass has a strong cutoff at 3\msol, above which no exoplanets have been found. Could it be that planets do not form around intermediate-mass stars, or is this a selection effect? The planet formation models of \cite{kennedy} suggest the latter. They predict that the fraction of stars with giant planets shows a steady increase with mass up to 3\msol. In addition, the mass of the planets and the width of
the regions where they form are predicted to increase with stellar mass. Probing intermediate-mass stellar systems for planets could help test these planet formation models.
Since most exoplanet detection methods tend to break down in the intermediate-mass range, one can look at these systems in post main-sequence where the contrast between the planet's photometric signal and that of its host is greatly decreased. This work examines a sample of white dwarfs decedent from $1-7M_{\odot}$ stars and constrains the frequency of planetary systems in the elusive intermediate-mass regime.

If planets survive post-MS evolution, they should be detectable around WDs \citep{burleigh,kilicgould}. Unfortunately, there are still no confirmed planets around single WDs. The candidate planet around the pulsating WD GD 66 \citep{mullally} is currently disputed (J. J. Hermes 2012, private communication). Perhaps an easier way to detect remnant planetary systems around WDs is to look for the tidally disrupted remains of exoplanets and moons in the form of circumstellar debris disks \citep{debes02, jura03, kilic06}. There are now two dozen WDs known to host dust and/or gas disks. Photospheric abundance analyses of these WDs show that the accreted metals originate from tidally disrupted minor bodies similar in composition to that of bulk Earth \citep{zuckerman,klein,klein11,dufour,dufourkilic}. Since at least one giant planet is required to perturb minor bodies out of their stable orbits \citep{debes}, photospheric pollution as well as circumstellar
debris disks serve as tracers for giant planets at WDs. Therefore, the frequency of debris disks around WDs can be taken as a lower
limit on the frequency of planets around WD remnants of intermediate-mass stars.

The search for debris disks around WDs begins as a search for excess emission in the infrared. Over the last two decades, several hundreds of WDs have been surveyed for infrared excess due to substellar companions or circumstellar disks. The first dusty WD, G29-38, was identified by \citet{zuckermanbecklin}, see also \citet{graham}. \citet{zuckerman92} examined 200 WDs for excesses in the K-band, while \citet{farihi05} surveyed 371 WDs of all spectral types and constrained the frequency of substellar companions to less than 0.5\%. The discovery of the second dusty WD, GD 362, \citep{kilic05,becklin05} lead to the realization that dusty WDs may be found more readily by looking at metal-rich WDs.

Previous surveys for debris disks around WDs focused on metal-rich WDs since so far all known WDs with disks also show metal absorption lines in their optical and ultraviolet spectra; DAZ and DBZ spectral types. The Poynting-Robertson drag timescale is significantly shorter than the evolutionary timescales for WDs \citep{vonhippel,rafikov}, hence we expect every WD with dust/gas disks to accrete metals from these disks and appear metal-rich. \citet{zuckerman03,zuckerman10} find that about 30\% of DA and DB WDs show trace amounts of metals in their photospheres. However, only $\sim20-30$\% of DAZs, the ones with the highest accretion rates, host  dust \citep{kilic08,farihi09}. The source of metals for the other DAZs remains uknown, but it is likely that the majority of DAZs accrete many small asteroids that do not form a large debris disk \citep{jura08}.

\citet{mullally} survey 124 WDs with $T_{\rm eff}=5,000-170,000$K using mid-infrared \emph{Spitzer} photometry. Their search for infrared excess finds 2 dusty WDs for a disk frequency of 1.6\%. However, this survey extends beyond the temperature range within which solid dust orbiting interior to the tidal radius can persist. Based on {\em Spitzer} observations of mostly metal-rich WDs, \citet{farihi09} estimate a debris disk frequency of 1\% to 3\% for WDs younger than 500Myr. \cite{debes11} recently searched for infrared excesses amongst an unbiased (in terms of metallicity) WD sample using the Wide-field Infrared Survey Explorer \citep[WISE,][]{wright10} data. They found about $1-5$\% of these WDs to be dusty. However, the large beam size of WISE allows for many false positives due to background contamination, and many of these dusty WD candidates need follow-up higher spatial resolution infrared imaging data to confirm the observed infrared excess. 

Here we present a near- and mid-infrared photometric and spectroscopic survey of an unbiased (in terms of metallicity) sample of DA WDs from
the Palomar-Green (PG) survey. Our survey does not focus on metal-rich WDs and thus constrains the disk frequency representative of the general WD population.
We target 117 PG WDs with accurate temperature, mass, and age estimates \citep{liebert05} and  $T_{\rm eff}=9,500-22,500$K. With 117 objects we sample temperatures at which solid dust can persist within the tidal radius and thus improve upon the constraint obtained by \citet{mullally} whose sample includes 69 WDs in this temperature range. The WDs in our sample are the descendants of MS stars with masses between $1 - 7$\msol, so this study extends the search for planets well beyond the mass range available to conventional exoplanet detection methods.  
Section 2 describes our near- and mid-infrared photometry and spectroscopy data, while Section 3 presents the spectral energy distributions
(SEDs) and the dusty WDs in the PG survey. Section 4 presents a discussion of the frequency of disks and planets around WDs and their progenitor
MS stars.
%
%
\section{OBSERVATIONS}
\subsection{The Sample}

Almost all known dusty WDs are relatively young with cooling ages of $\lesssim 1$Gyr and temperatures in the range $9,500-22,500$K
(with the exception of G166-58, Farihi et al. 2008). This is probably because a second heavy bombardment phase occurs around intermediate-mass stars in their post-MS evolution due to planetary migration around the mass-losing star \citep{debes, bonsorwyatt12}. We would expect to
see a peak in the frequency of collisions, tidal disruptions, and disks around the younger WD remnants of such stars \citep{debes02}.

Consider the disk around G29-38 as a typical system. \citet{jura03} models the disk emission and obtains a ring of dust extending from $R_{\textrm{\tiny in}}=0.1 R_{\odot}$ to $R_{\textrm{\tiny out}}=0.4-0.9 R_{\odot}$ and $T_{\rm eff}=300-600$K. Such a disk around a star twice or three times as hot as G29-38 would be sublimated and invisible in the IR. Thats why our target selection using effective temperature gives a better measurement of the real frequency of disks.

We select 117 apparently single DA WDs from the PG survey with $T_{\rm eff} = 9,500-22,500$K, where we are most efficient in finding the disks
using near- and mid-infrared data. \citet{liebert05} performed a detailed model atmosphere analysis of all DA WDs in the PG survey
and provided temperature, surface gravity, mass, and age estimates. Their spectroscopy did not have enough
resolution to detect the metal-rich DAZs among these targets; our survey does not have a metallicity bias, and hence it has the potential to
reveal the true frequency of dusty disks without knowing the fraction of metal-rich WDs among the nearby WD population.

\subsection{Near-infrared Photometry}

We obtained simultaneous \emph{JH$K_s$} imaging of 78 WDs using the Peters Automated Imaging Telescope \citep[PAIRITEL,][]{bloom} between 2009
Feb and 2010 Apr. PAIRITEL is the old Two Micron All Sky Survey \citep[2MASS,][]{cutri03} telescope and it uses the same camera and filter set.
The total exposure time for our targets ranged from 70 s to 1,500 s depending on the expected brightness of each target. We use the PAIRITEL pipeline version 3.3 processed images and standard IRAF DAOPHOT routines to perform aperture photometry on every 2MASS source detected in the images. We use the 2MASS stars to calibrate the photometry for each object.

\begin{figure}
	\centering
	\includegraphics[width=\textwidth]{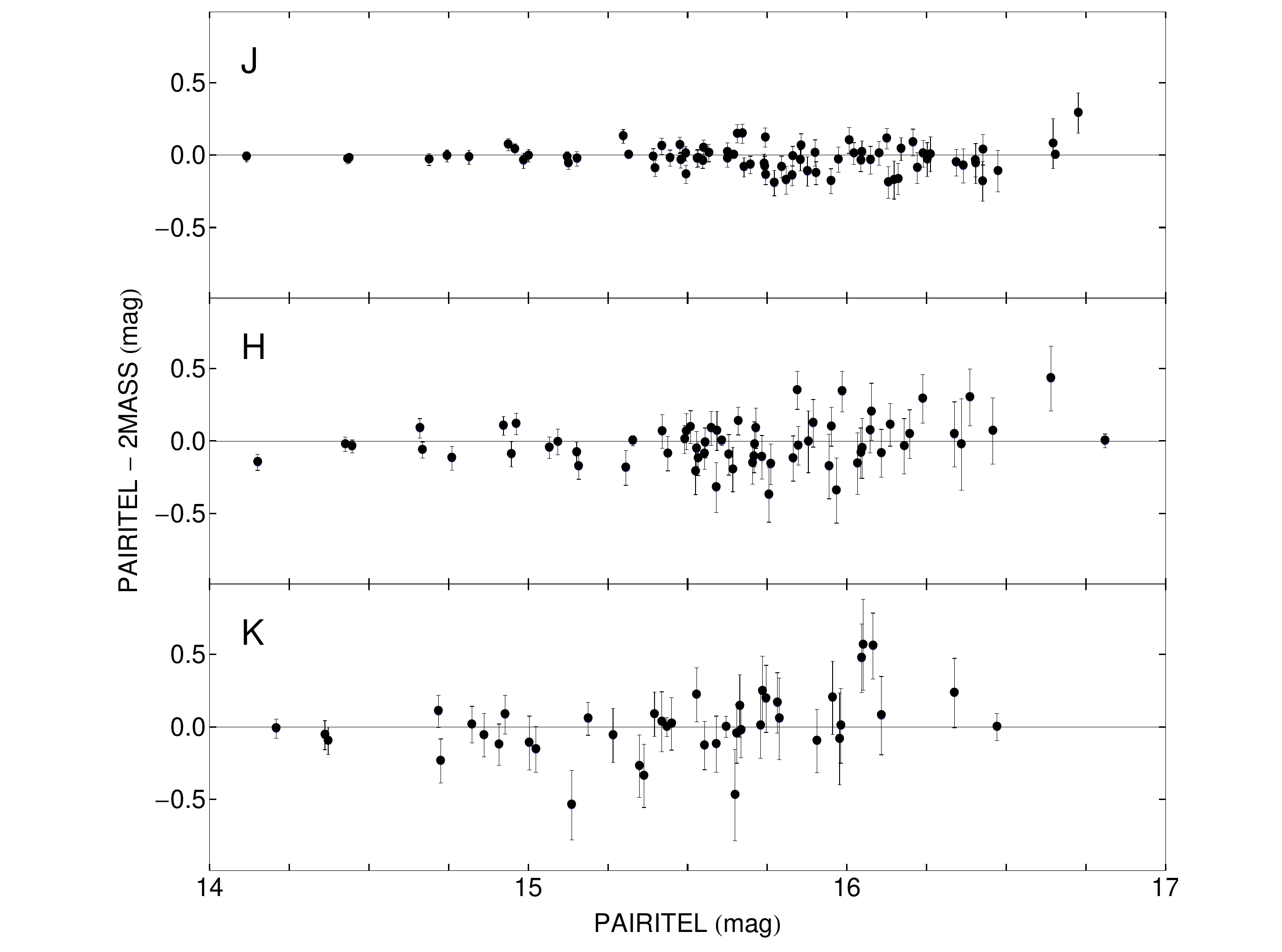}
	\caption{The PAIRITEL vs. 2MASS photometry for our sample. The PAIRITEL data agrees fairly well with the 2MASS photometry.}
	\label{pairitel}
\end{figure}

Table 1 presents the PAIRITEL photometry (and the 2MASS photometry of a few relatively bright targets) for our sample along with the
physical parameters obtained from the model atmosphere analysis of \citet{liebert05}. 
We compare our PAIRITEL photometry to that of 2MASS in Figure \ref{pairitel}. Only about half of our targets are detected by 2MASS in the K-band,
and usually with large photometric errors. Our survey goes significantly deeper than 2MASS, and provides improved photometry with smaller
errors, which is essential for identifying objects with slight $K-$band excesses. This figure demonstrates that
the PAIRITEL photometry agrees fairly well with the 2MASS photometry and that it can be used reliably to constrain the near-infrared SEDs
of our targets. 

\begin{figure}
	\centering
	\includegraphics[width=\textwidth]{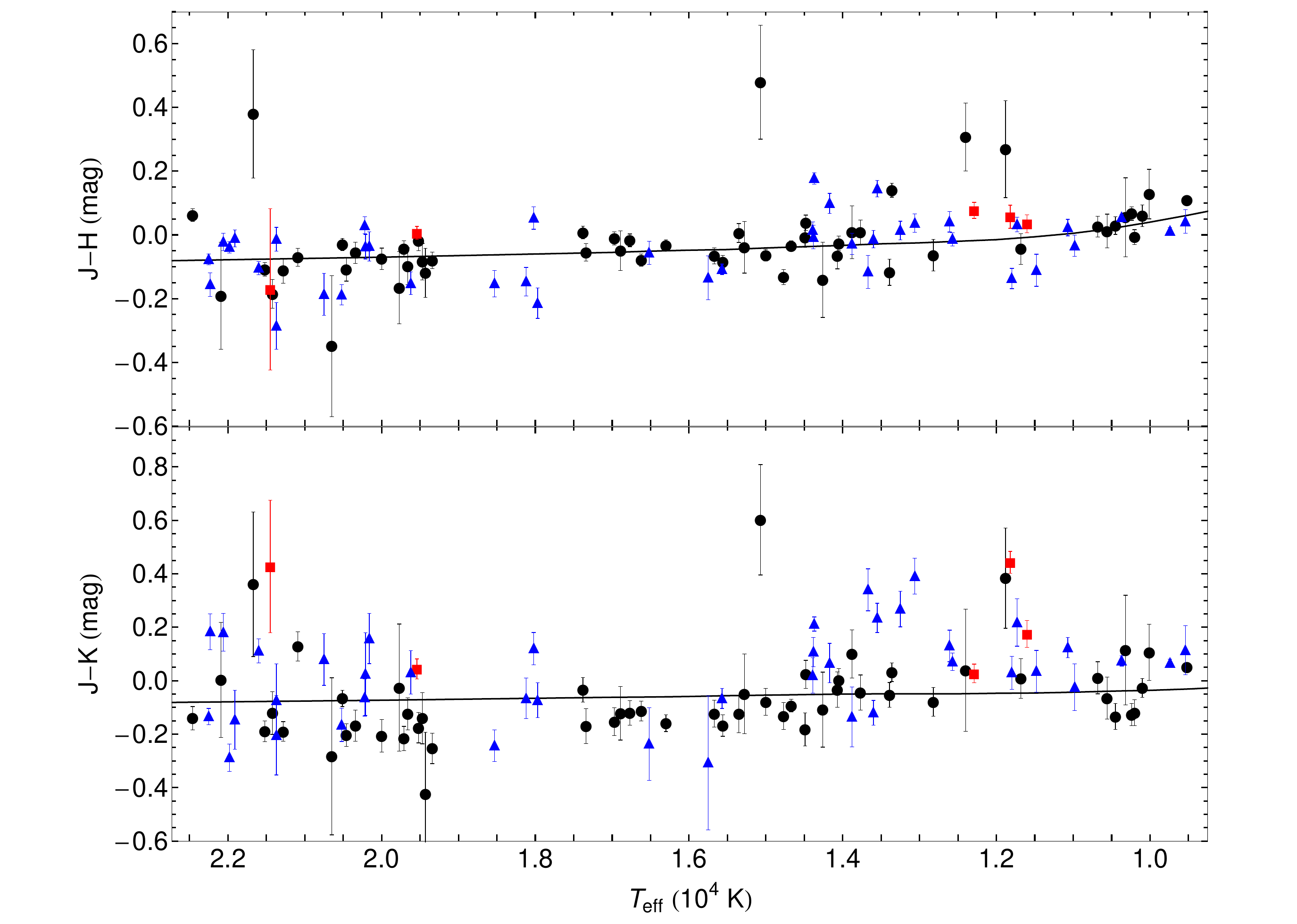}
	\caption{$J-H$ and $J-K$ colors vs. temperature for our sample of DA WDs from the PG survey. Red squares and blue triangles show the confirmed dusty WDs and objects with follow-up IRTF spectroscopy, respectively. Solid lines show the predicted sequences for DA WDs \citep{bergeron}.}
	\label{JHK}
\end{figure}

Figure \ref{JHK} shows $J-H$ and $J-K$ colors of our sample of stars as a function of their effective temperatures. The colors for the majority
of our targets follow the model predictions of $J-H = J-K \approx 0$ mag, though with some scatter. There are four WDs with $J-H$ excesses, though
with relatively noisy 2MASS photometry. However, from the available mid-infrared photometry from \emph{Spitzer} and/or WISE, we conclude that there is no evidence of excess in these objects. The $J-H$ photometry for the rest of the targets are consistent with the emission from single WDs. Several of these apparently single WDs have positive $J-K$ colors,
suggesting extra emission from a cool companion or a circumstellar debris disk. We chose to further examine those objects with positive $J-K$
colors by obtaining near-infrared spectra and mid-infrared photometry. Objects with apparently red colors were targeted however, depending on the timing of the observing runs and conditions, some of the objects with blue colors were also followed up as a sanity check.

\subsection{Near-infrared Spectroscopy}

We obtained low-resolution near-infrared spectra of 41 WDs over several nights in 2011 April, August, and September using the 3-m NASA
InfraRed Telescope Facility (IRTF) equipped with the $0.8-5.4$ Micron Medium-Resolution Spectrograph and Imager (Spex; \citealt{rayner}).
The observing setup and procedures are similar to those of \cite{barber}. We use a $0.5\arcsec$ slit to obtain a resolving power of 90-210
over the $0.7-2.5\micron$ range. The observations are taken in two different positions on the slit separated by $10\arcsec$. We use internal
calibration lamps (a 0.1W incandescent lamp and an Argon lamp) for flat-fielding and wavelength calibration, respectively. To correct for
telluric features and flux calibrate the spectra, we use the observations of nearby A0V stars at a similar airmass to the target observations.
We use the IDL-based package SPEXTOOL version 3.4 (\citealt{cushing}) to reduce the data.

\begin{figure}
        \centering
        \includegraphics[width=\textwidth,trim=.8cm 0cm 0cm 0cm,clip=true]{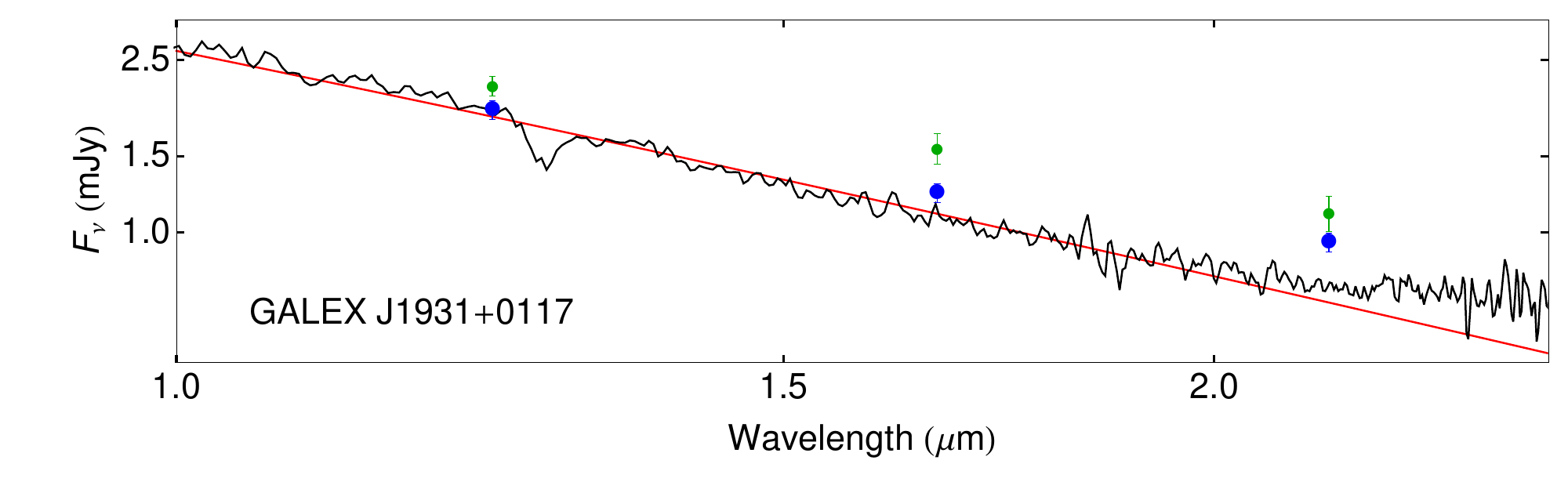}
        \caption{Spectral energy distribution of the known dusty WD GALEX J1931+0117. Green points represent 2MASS photometry, while blue points represent the photometry from \citealt{melis11}. The IRTF spectrum is shown in black and the red line represents the predicted contribution from a blackbody stellar photosphere. The noise near 1.9 and 2.5 $\mu$m falls in regions of poor telluric correction. The 1.28 $\mu$m feature is Pa$\beta$ absorption. The detection of slight K-band excess around this target demonstrates that our near-infrared observations and reductions are reliable.}
        \label{GALEX}
\end{figure}

To demonstrate that our near-infrared observations and reductions are reliable, we also obtained a near-infrared spectrum of the previously
known dusty WD GALEX J193156.8+011745 \citep[hereafter J1931+0117;][]{vennes10,debes11,melis11}. Figure \ref{GALEX} shows the near-infrared photometry and our IRTF spectrum
of this object. There are two nearby red sources within 2$\arcsec$ of the WD \citep{melis11}. Our IRTF observations were
obtained under average seeing conditions of 0.8$\arcsec$ and our 0.5$\arcsec$ slit minimizes the contamination from the nearby sources.
A slight excess is clearly detected redward of 2.1$\micron$, typical of dusty WDs \citep{kilic06}. \citet{melis11} also obtained
a near-infrared spectrum of J1931+0117 on the Magellan 6.5m Baade telescope, which shows a significantly larger excess redward of 1.6$\micron$.
Those authors discuss the potential problems with the flux calibration of their spectra. They model the excess around J1931+0117 with
a flat disk model with inner and outer temperatures of 1400 and 1200 K, essentially a narrow ring of dust. Our IRTF spectrum demonstrates that 
the excess around J1931+0117 mostly shows up redward of 2.1$\micron$ and it is similar to the other dusty WDs with extended (up to 1$R_{\odot}$)
disks. Of course, the outer edge of the disk around J1931+0117 is currently unconstrained due to the lack of uncontaminated mid-infrared photometry.
Regardless of these issues, this figure demonstrates that we are able to identify dusty WDs even with slight $K-$band excesses.

\subsection{\emph{Spitzer} Photometry}

We used the warm {\em Spitzer} equipped with the InfraRed Array Camera (IRAC; \citealt{fazio}) to obtain infrared photometry of 11 WDs between
2010 Sep and 2011 Feb for program number 70023. Based on our preliminary analysis of the PAIRITEL data, these 11 sources appeared to have slight
$K-$band excesses indicative of debris disks. We obtained 3.6 and 4.5$\micron$ images with integration times of 30 s or 100 s for nine dither
positions. We use the IDL ASTROLIB packages to perform aperture photometry on the individual corrected basic calibrated data frames from the S18.18.0 pipeline reduction.

Following the IRAC calibration procedures, we correct for the location of the source in the array before averaging the fluxes of each of the dithered frames at each wavelength. We also correct the Channel 1 (3.6\micron) photometry for the pixel phase dependence. We estimate the photometry error bars from the observed scatter in the nine images corresponding to the dither positions. We add the 3\% absolute calibration error in quadrature \citep{reachb}. Finally, we divide the estimated fluxes by the color corrections for a Rayleigh-Jeans spectrum, except for the dusty WDs in our
sample. Table \ref{table2} presents our {\em Spitzer} IRAC photometry for 11 PG WDs. 
%
%
\section{RESULTS}
\begin{figure}
	\centering
	\includegraphics[width=.45\textwidth,trim=7cm 0cm 7cm 0cm,clip=true]{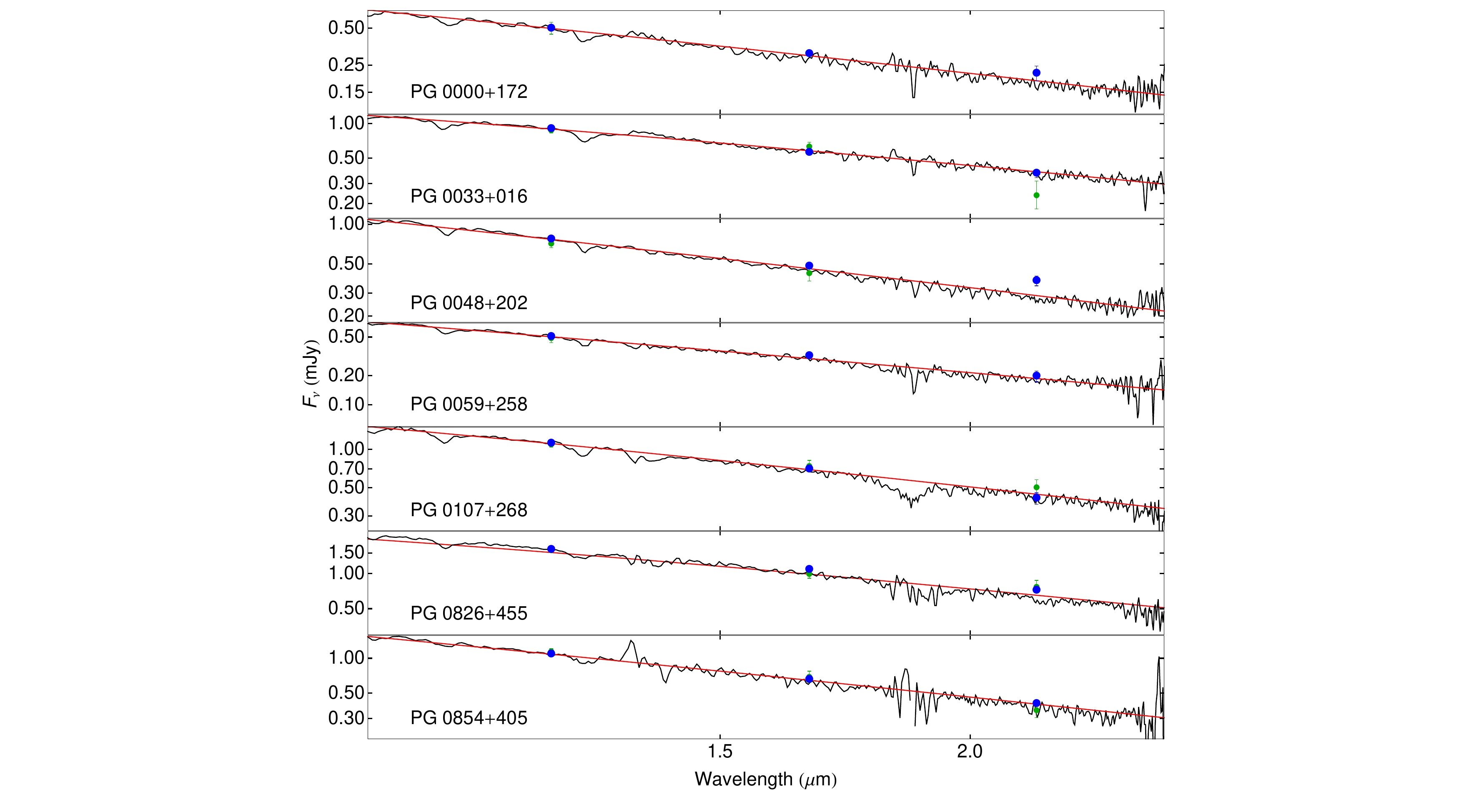}
	\includegraphics[width=.45\textwidth,trim=7cm 0cm 7cm 0cm,clip=true]{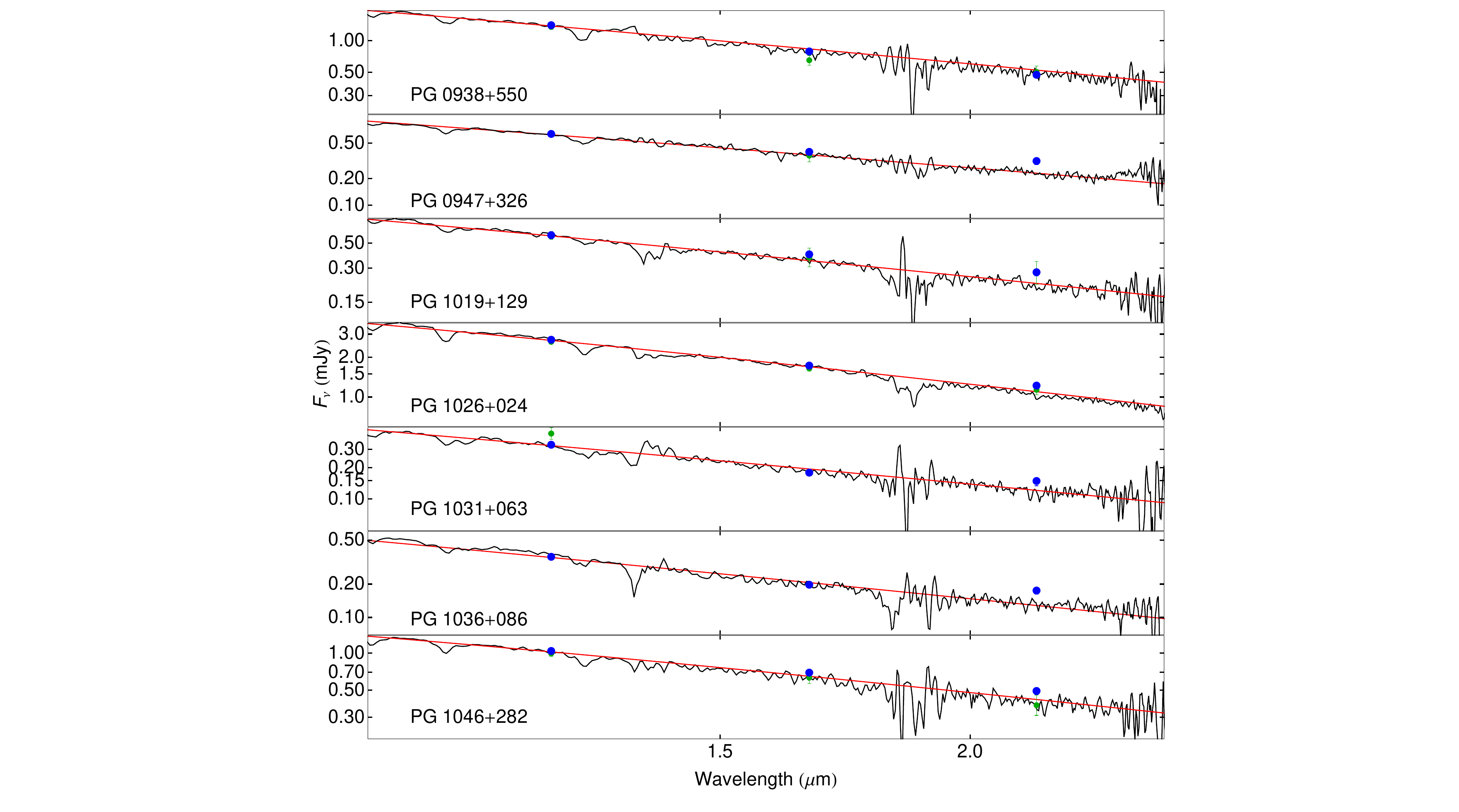}
	\includegraphics[width=.45\textwidth,trim=7cm 0cm 7cm 0cm,clip=true]{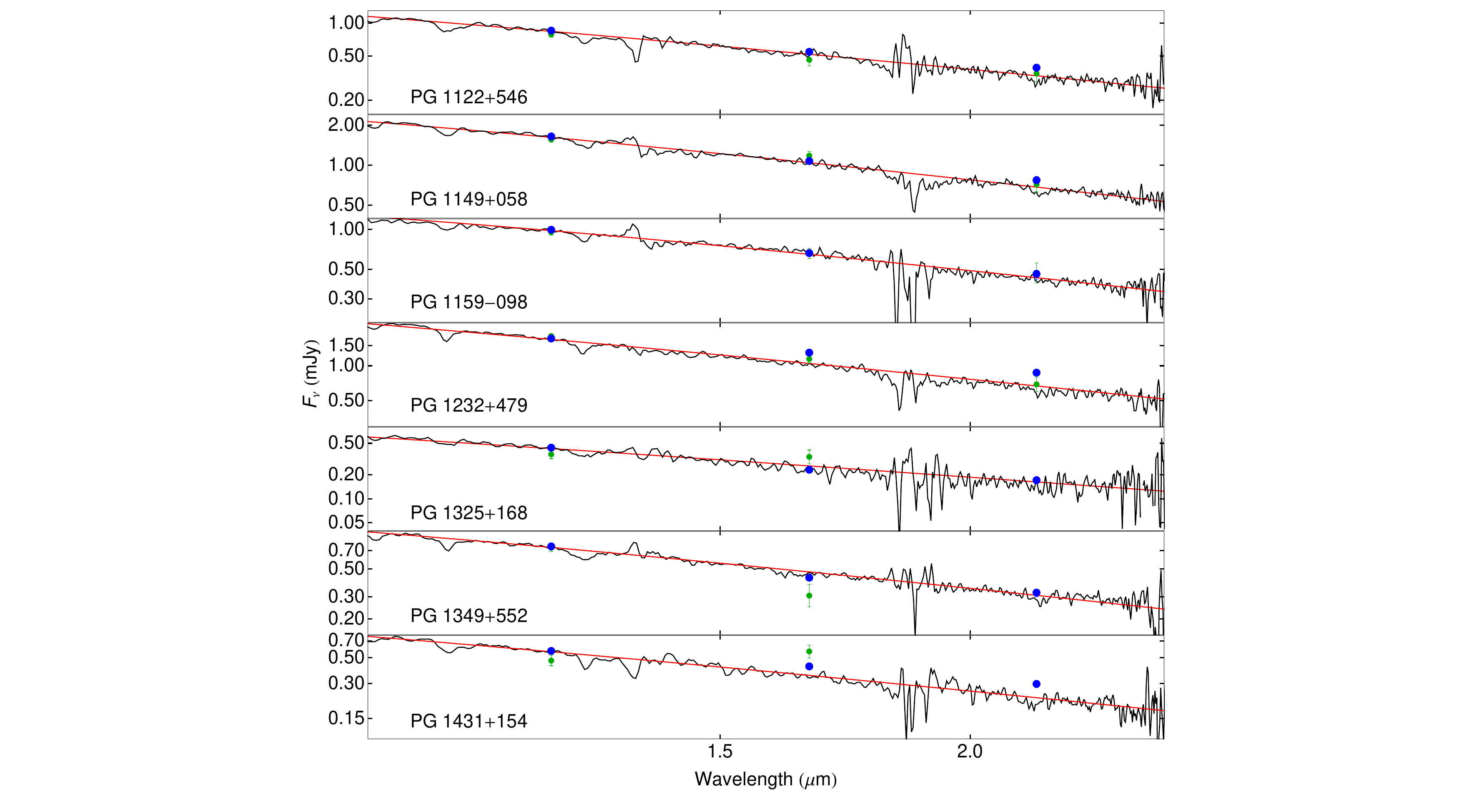}
	\includegraphics[width=.45\textwidth,trim=7cm 0cm 7cm 0cm,clip=true]{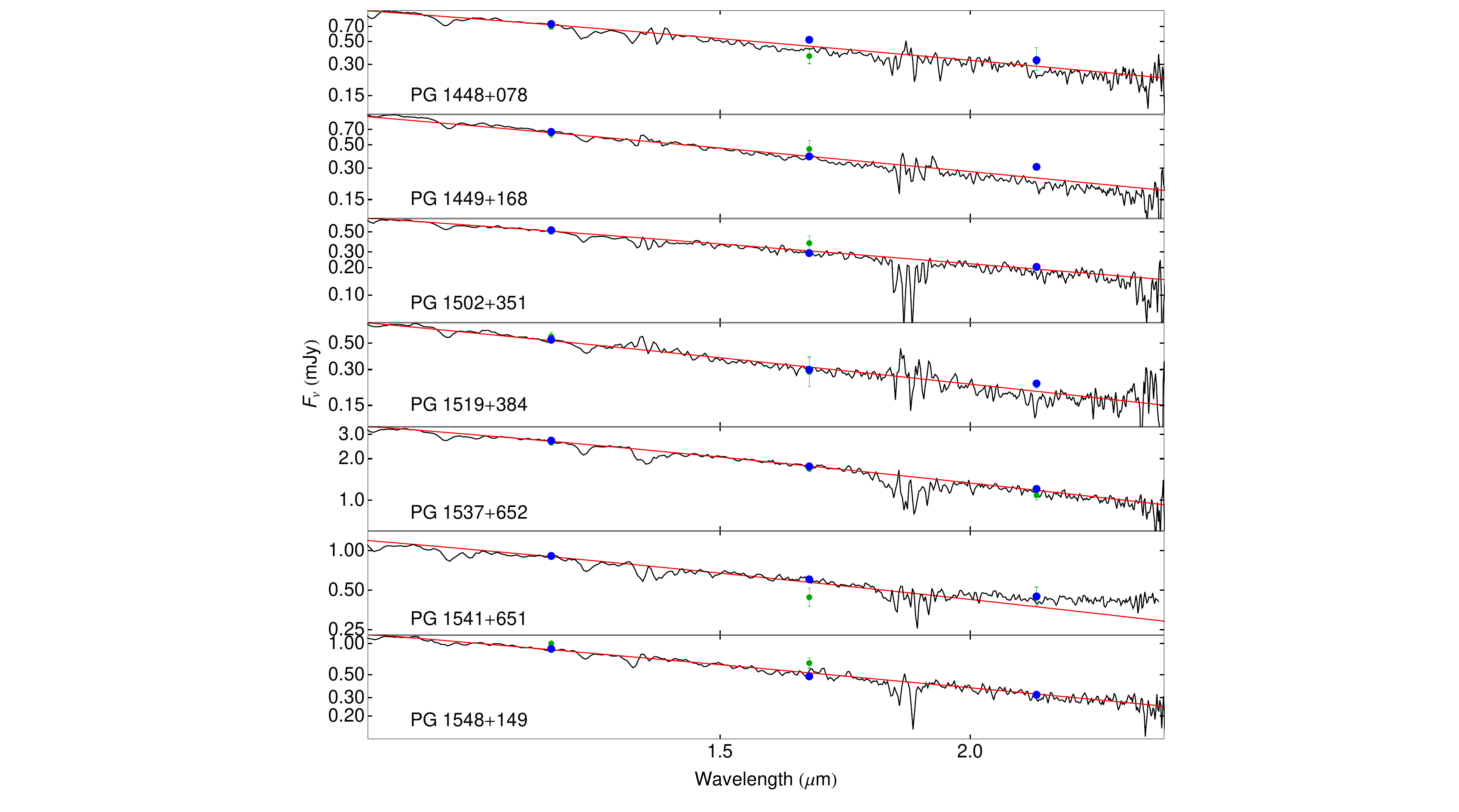}
	\includegraphics[width=.45\textwidth,trim=7cm 0cm 7cm 0cm,clip=true]{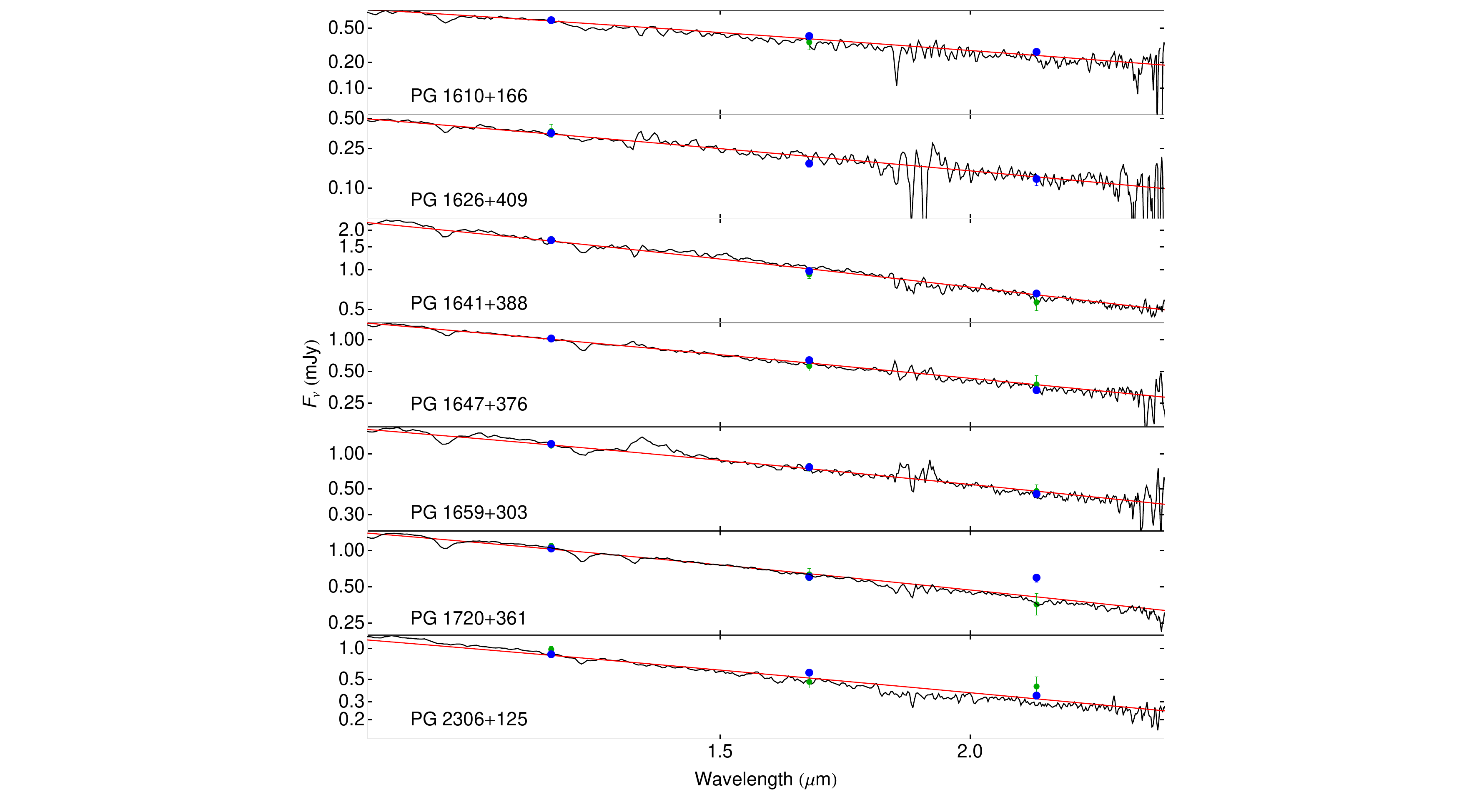}
	\includegraphics[width=.45\textwidth,trim=7cm 0cm 7cm 0cm,clip=true]{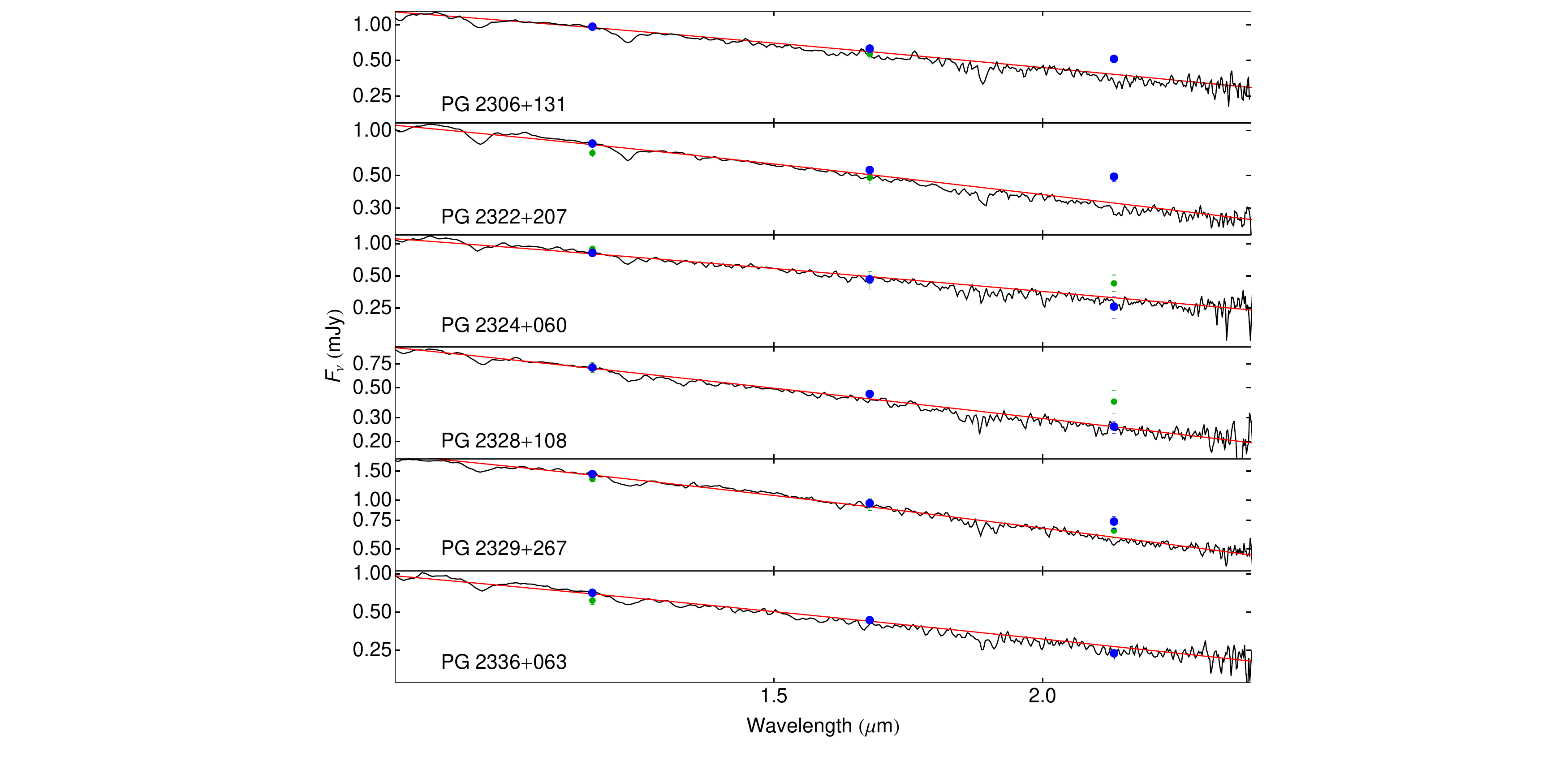}
	\caption{Spectral energy distributions of 41 WDs observed at the IRTF. The PAIRITEL and 2MASS photometry are shown as blue and green points, respectively. The black lines show the observed spectra and the red solid lines show the predicted photospheric emission for each star assuming a blackbody. The IRTF spectra display telluric correction problems around 1.4 and 1.9 $\mu$m.}
	\label{irtf}
\end{figure}

Our near-infrared photometric observations revealed several sources with potential $K-$band excesses. We followed up 11 of the most interesting
targets with {\em Spitzer} IRAC and 41 stars in total at the IRTF. Figure \ref{irtf} shows the SEDs of the 41 stars observed at the IRTF. Here, and
in the rest of the figures, the PAIRITEL and 2MASS photometry are shown as blue and green points, respectively. The black lines show the
observed spectra and the red solid lines show the predicted photospheric emission for each star assuming a blackbody. The IRTF spectra
display telluric correction problems around 1.4 and 1.9$\micron$, but otherwise they closely follow the predicted blackbody distributions
for the majority of the targets in our sample. PG 0048+202 and PG 1720+361 are two of the stars for which our PAIRITEL photometry indicated
$K-$band excesses, but our IRTF spectra show that there are no significant excesses in the $K-$band for these two stars. There is only one
WD in our IRTF sample where a significant $K-$band excess is detected, PG 1541+651. The observed flux stays essentially constant between 2.0 and 2.5 $\micron$, which is typical of dusty WDs. The detection of Ca II 3933 \AA~ absorption in the HIRES spectrum of this object (Xu et al. in preparation) corroborates the presence of circumstellar debris.

\begin{figure}
	\centering
	\includegraphics[width=.85\textwidth,trim=7cm 0cm 7cm 0cm,clip=true]{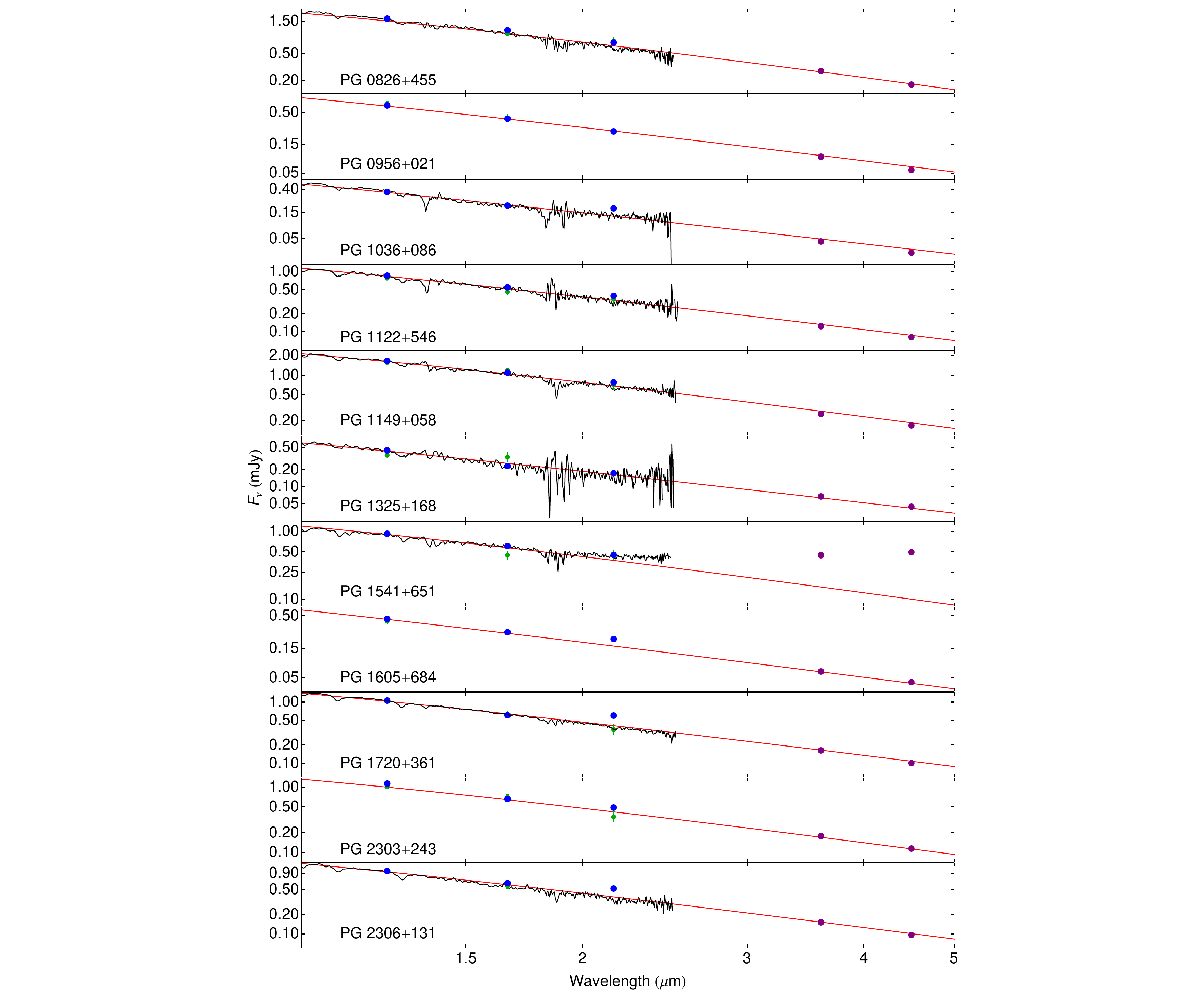}
	\caption{SEDS of the 11 WDs with new \emph{Spitzer} photometry (shown in purple). The symbols are the same as in Figure \ref{GALEX}. The IRTF spectra display telluric correction problems around 1.4 and 1.9 $\mu$m. Our \emph{Spitzer} observations confirm the results of the IRTF observations; only PG1541+651 shows a significant infrared excess.}
	\label{SpitzerCycle7}
\end{figure}

Figure \ref{SpitzerCycle7} shows the near- and mid-infrared data on 11 WDs that we targeted with {\em Spitzer}, including PG 1541+651.
Eight of these stars also have IRTF spectroscopy available. Our {\em Spitzer} observations confirm the results from the IRTF observations;
only PG 1541+651 shows a significant infrared excess that is compatible with a debris disk \citep[see][]{barber}. The SEDs for the other
ten objects in this figure are consistent with photospheric emission from each star.

\begin{figure}
	\centering
	\includegraphics[width=\textwidth,trim=7cm 0cm 7cm 0cm, clip=true]{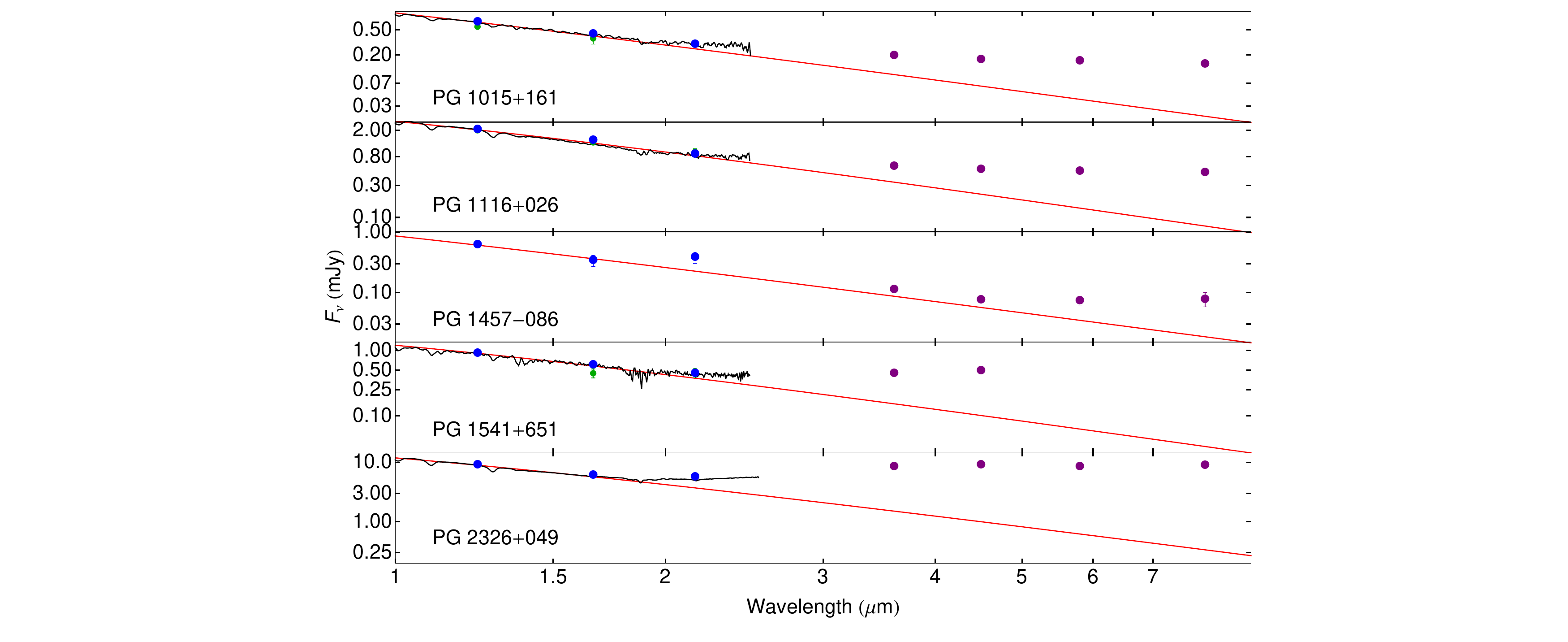}
	\caption{SEDs of the five dusty WDs in our sample. The symbols are the same as in Figure \ref{SpitzerCycle7}. The IRTF spectra display telluric correction problems around 1.4 and 1.9 $\mu$m.}
	\label{Disks}
\end{figure}

There are four other previously known dusty WDs in our sample; PG 1015+161, PG 1116+026, PG 1457$-$086, and PG 2326+049. Figure \ref{Disks} displays the SEDs for these four WDs plus PG 1541+651. PG 2326+049 has the most prominent excess, while the excess at PG 1457$-$086 is the most subtle.
The variation in the strength of these excesses is likely due to a variation in the disk geometry from object to object.
There is another known dusty WD in the \citet{liebert05} DA WD sample, PG 1456+298 \citep[G166-58,][]{farihi} that is cooler
than the lower temperature limit of our sample and hence it is not included in Figure \ref{Disks}.

\begin{figure}
	\centering
	\includegraphics[width=.75\textwidth]{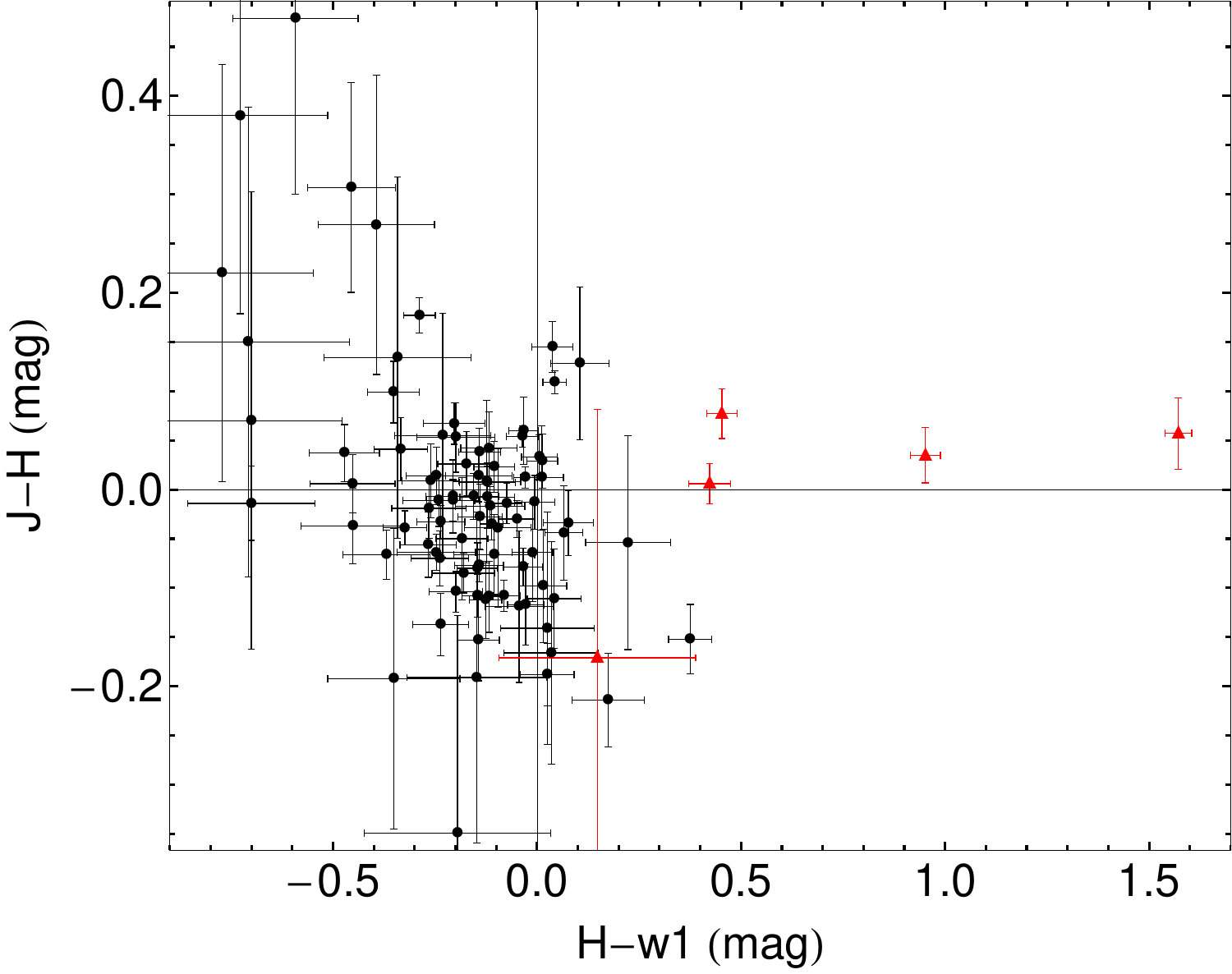}\\~\\
	\includegraphics[width=.75\textwidth]{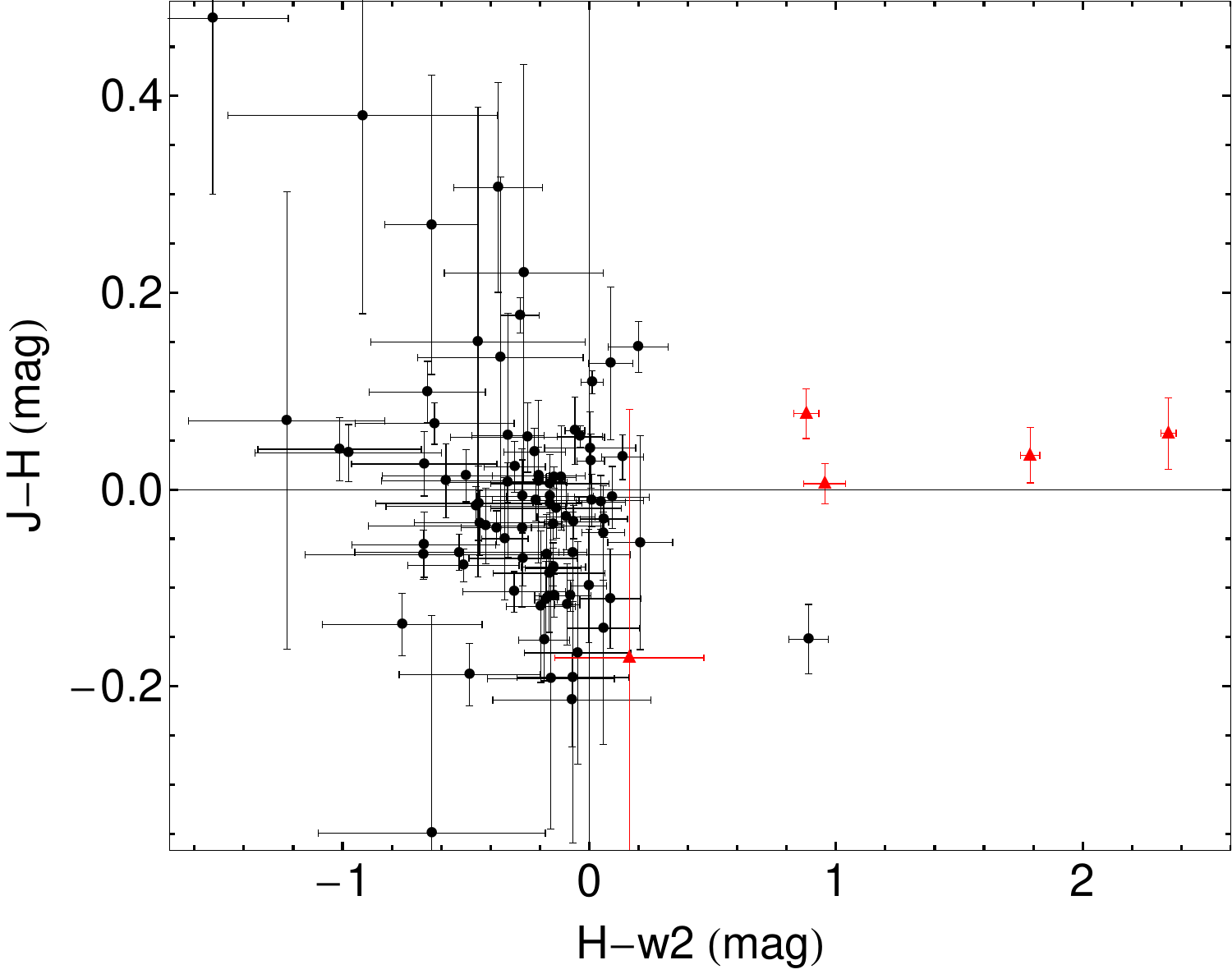}
	\caption{Color-color diagrams using PAIRITEL/2MASS and WISE photometry for our WD sample. Red triangles mark the dusty WDs. Four of the five dusty WDs in our sample clearly show infrared excesses in the WISE bands. The fifth dusty WD, PG 1457$-$086, is also detected in the WISE observations, but with large photometric errors.}
	\label{WISE}
\end{figure}

Out of the 117 DA WDs in our sample, the majority have SEDs that closely follow a single blackbody curve, while five WDs show a
significant deviation from their blackbody model in the near- to mid-infrared (see Figure \ref{Disks}).
In light of the recent WISE All-Sky Data Release, we examine our sample of objects for infrared excesses in the WISE data and plot
these results in Figure \ref{WISE}. Four of the five dusty WDs in our sample clearly show infrared excesses in the WISE bands. The fifth dusty WD, PG 1457$-$086, is also detected in the WISE observations, but with large photometric errors. There is another target, PG 1519+384, with positive $H-W1$ and $H-W2$ colors. However, there are two nearby sources within several arcseconds of this target in the Sloan Digital Sky Survey images. The WISE photometry for PG 1519+384 is likely contaminated by these background sources. Hence, the WISE data do not reveal any new dusty WDs in our sample other than the five systems known.

 \citet{farihi05} finds a frequency of less than 0.5\% for brown dwarf companions of WDs. Given our sample size of 117 we would expect to find less than one brown dwarf companion in our sample. It is not surprising, then, that we do not find any substellar companions. Through the initial selection process, we excluded objects with potential M-dwarf companions.
\section{DISCUSSION}
\subsection{The Frequency of Debris Disks Around WDs}

Out of the 348 DA WDs analyzed by \citet{liebert05}, 308 are apparently single. Six of these PG WDs are known to host dust disks, corresponding to a debris disk frequency of 1.9\%. This is similar to the frequency of disks, 1.6\%, around the nearby bright WD sample of \citet{mullally}. Of course, these estimates ignore the fact that most disks occur around young WDs
\citep[see][]{kilic09}. The process that creates the debris disks around WDs seem to be more efficient at younger ages.

\begin{figure}
	\centering
	\includegraphics[width=0.95\textwidth]{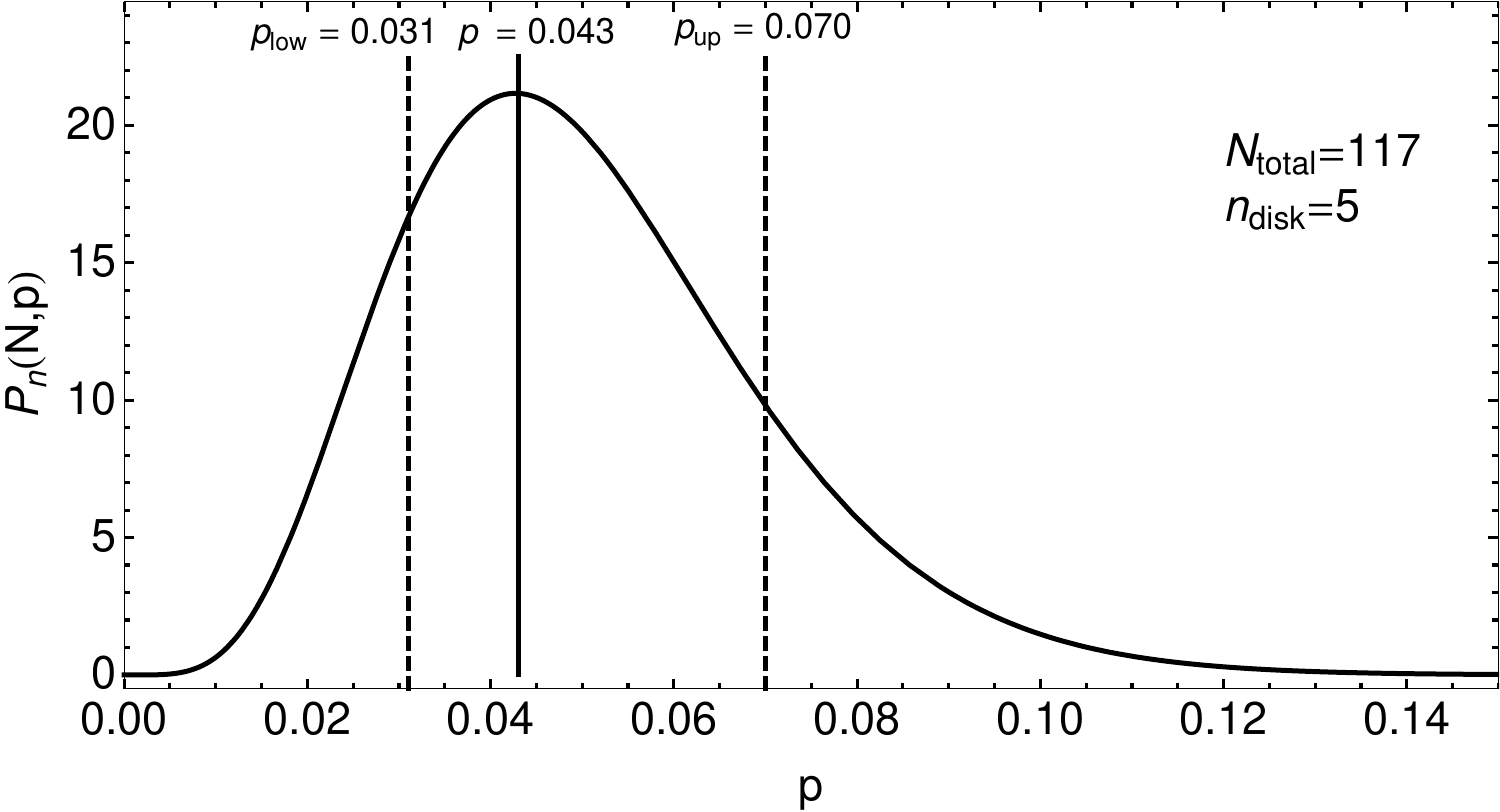}
	\caption{Probability function for WDs with debris disks. Probability, $P_n(N,p)$, of finding $n=5$ debris disks as a function of assumed true debris disk frequencies, $p$, where $N=117$. Dashed lines delineate the region containing 68\% probability, equivalent to 1$\sigma$ Gaussian limits. The solid line indicates the peak of the distribution. The derived frequency is $4.3^{+2.7}_{-1.2}\%.$}
	\label{probability}
\end{figure}

We use the binomial probability distribution to compute the upper and lower limits on the frequency ($p$) of disks. The binomial
distribution function gives the discrete probability distribution of obtaining exactly $n$ successes out of $N$ trials (where
the result of each trial is true with probability $p$ and false with probability $1-p$). The probability, $P_n(N,p)$, that a survey
of $N$ WDs will detect $n$ debris disks, when the true frequency of disks is $p$ is given by \citep[Appendix]{burgasser} 
		\begin{equation}
			P_n(N,p)=\left(\begin{array}{c}N\\n\end{array}\right)p^n(1-p)^{N-n}=\frac{N!}{n!(N-n)!}p^n(1-p)^{N-n}.
		\end{equation}
Since this probability function is not symmetric about its maximum value, we report the range in $p$ that delimits 68\% of the integrated probability function, equivalent to 1$\sigma$ Gaussian limits. For $N=117$ and $n=5$, there is a 68\% chance of $p$ being between 3.1\% and 7\% (see Figure \ref{probability}). Thus, $p=4.3^{+2.7}_{-1.2}\%$.
This frequency estimate is slightly higher than, but consistent with, the 3\% disk frequency derived by \citet{farihi09}.

\subsection{Planets Around Intermediate-Mass Stars}		

The 117 WDs in our sample have an average mass of $0.6M_{\odot}$ and mass range of $0.4-1.2M_{\odot}$.
We estimate the progenitor MS masses using the initial-final mass relation derived by \citet{kalirai} and \citet{williams}.
We obtain an average progenitor mass of $2.4M_{\odot}$ and mass range of $1.1-7.2M_{\odot}$ (see Table \ref{table1}) where
we neglect WDs below 0.5$M_{\odot}$. Based on the observed frequency of disks around our WD targets,
we find that the probability of finding planetary systems around their progenitor $1-7M_{\odot}$ MS stars is at least \freq.
This result serves as an indirect probe of the intermediate-mass stellar systems which have thus far challenged conventional
planetary detection techniques \citep{kilicgould}.

The dusty WDs in our sample, PG 1015+161, 1116+026, 1457$-$086, 1541+651, and 2326+049 are the descendants of $2.1-2.8M_{\odot}$ MS stars.
At first glance, this argues that planet formation may be most efficient around $2-3M_{\odot}$ stars. However, given the age and star formation
history of the Galactic disk, our sample is dominated by the $\sim 0.6M_{\odot}$ WD remnants of $2-3M_{\odot}$ MS stars. In fact, 80\%
of our sample consists of WD descendants of $1-3M_{\odot}$ MS stars. Hence, the lack of discoveries of dusty disks, and therefore remnant
planetary systems, around the massive WD remnants of $M >3M_{\odot}$ stars may be due to small number statistics. There are 22 systems
in our sample with $M >3M_{\odot}$ progenitors, we would expect to find 1 dusty WD in a sample of 22 stars. The probability of finding zero dusty WDs in a sample of 22, when the expected number is one, is 30\% (see Equation 1).
Hence, larger surveys of massive WDs are required to detect disks around them and to constrain the frequency of planets around
their intermediate-mass progenitor MS stars.

\subsection{Disk Lifetimes}

An alternative interpretation of the 4.3\% disk frequency around WDs is that perhaps 100\% of WDs have remnant planetary systems that can
create debris disks, but the disks only persist for 4.3\% of the cooling age of the WD.
For the average cooling age of our sample of WDs, 300Myr, this implies that disks survive around typical WDs for $\sim$10Myr.
We note that this is the total time a WD accretes from circumstellar debris disks, even though individual disks may last for a significantly
shorter amount of time. \citet{kilic08} and \citet{farihi09} estimate typical disk lifetimes of $\sim10^5$ yr. Hence, up to 100
tidal disruption events that create circumstellar debris disks may occur for any given WD.

\citet{vonhippel} use a geometrically thin, optically thick disk model to determine typical accretion rates of WDs. Their result ($\sim 10^9$ g s$^{-1}$) means a WD with the disk lifetime typical of our sample will accrete a total mass equivalent to a 400 km Solar System asteroid (assuming an asteroid density of 3 g cm$^{-3}$). \citet{rafikov11a} demonstrate that the Poynting-Robertson drag can explain accretion rates of around $10^8$ g s$^{-1}$ from circumstellar debris disks onto the WDs. Taken at face value, this implies a total accreted mass of $10^{22}$ g, a 200km asteroid, over the disk-hosting lifetime of a WD. However, dusty WDs show accretion rates of up to $10^{11}$ g s$^{-1}$ \citep{dufourkilic}. To explain the observed high accretion rates,
\citet{rafikov} and \citet{metzger12} propose a runaway accretion scenario where rapid transport of metals from the disk results from
interaction between spatially coexisting dust and gas disks. In this scenario, a WD accretes $10^{22}$ g of metals in $10^5$ years.
Hence, typical WDs may accrete up to $10^{24}$ g of metal over 10Myr, the total time they host circumstellar disks. This is comparable to the mass of the dwarf planet Ceres and Pluto's moon Charon. Of course, if the tidally disrupted object is icy, the total accreted mass may be significantly higher \citep[although see][]{jura12}. Therefore, the debris surrounding WDs is likely supplied through disruption of Solar System asteroid analogues, moons, and dwarf planets. 
\section{CONCLUSIONS}
We present near- and mid-infared observations and a comprehensive study of the SEDs of a metallicity-unbiased sample of 117 DA WDs from the
PG survey. Only five of our targets show excess radiation from debris disks, indicating a debris disk frequency of \freq~which fits in well with previous surveys of this kind.
We interpret this frequency as a lower limit to the frequency of planets around the $1-7M_{\odot}$ progenitor MS stars; an indirect
result for the intermediate-mass regime in which conventional exoplanetary detection methods are insensitive.
Alternatively, we interpret the observed frequency of disks as the fraction of time a WD accretes from its debris disks over the entire cooling
age of the star. We estimate that typical WDs may accrete metals from several generations of disks for $\sim$10Myr, corresponding to
accretion of metals up to $10^{24}$ g. This means WDs are capable of accreting bodies as large as dwarf planets as well as Solar System
asteroid analogues.

\acknowledgements

This work is based, in part, on observations made with the \emph{Spitzer Space Telescope}, which is operated by the Jet Propulsion Laboratory, California Institute of Technology under a contract with NASA. Support for this work was provided by NASA. The PAIRITEL is operated by the Smithsonian Astrophysical Observatory (SAO) and was made possible by a grant from the Harvard University Milton Fund, the camera loan from the University of Virginia, and the continued support of the SAO and UC Berkeley. The PAIRITEL project and JSB are further supported by NASA/Swift Guest Investigator Grant no.~NNG06GH50G. We thank M.~Skrutskie for his continued support of the PAIRITEL project. The IRTF is operated by the University of Hawaii under Cooperative Agreement no.~NNX-08AE38A with the National Aeronautics and Space Administration, Science Mission Directorate, Planetary Astronomy Program. We thank A.~Gianninas for his helpful comments, J. Farihi for useful discussions, and C.~Blake, M.~Skrutskie, and E.~Falco for helping with the PAIRITEL observations. We also thank the referee, Marc Kucher, for insightful suggestions.
%
%

%
%
\begin{deluxetable}{ccccccccccc}
\tabletypesize{\footnotesize}
\tablewidth{0pt} \rotate
\tablecaption{Sample Properties}
\tablehead{																																						
	&	\multicolumn{1}{c}{$T_{\textrm{eff}}$}	&		&	WD Mass	&	MS Mass	&	\multicolumn{1}{c}{$\tau$}	&	J			&	H			&	K			&											&				\\
PG	&	\multicolumn{1}{c}{(K)}	&	$\log g$	&	($M_{\odot}$)	&	($M_{\odot}$)	&	\multicolumn{1}{c}{(Myr)}	&		(mag)		&	(mag)			&		(mag)		&	Ref.										&		Data		
}																																						
\startdata																																						
0000+172	&	20210 	&	7.99 	&	0.62	&	2.1	&	85	&	16.262	$\pm$	0.016	&	16.299	$\pm$	0.035	&	16.237	$\pm$	0.154	&	1		   	   	   			   	   	   	&	I,		P	\\
0033+016	&	10980 	&	8.83 	&	1.12	&	6.7	&	2089	&	15.625	$\pm$	0.014	&	15.659	$\pm$	0.030	&	15.649	$\pm$	0.086	&	1		   	   	   			   	   	   	&	I,		P	\\
0048+202	&	20160 	&	7.99 	&	0.62	&	2.1	&	85	&	15.795	$\pm$	0.020	&	15.831	$\pm$	0.041	&	15.637	$\pm$	0.092	&	1										&	I,		P	\\
0059+258	&	21370 	&	8.04 	&	0.65	&	2.3	&	78	&	16.252	$\pm$	0.018	&	16.266	$\pm$	0.033	&	16.325	$\pm$	0.135	&	1										&	I,		P	\\
0107+268	&	13880 	&	7.87 	&	0.54	&	1.3	&	245	&	15.392	$\pm$	0.015	&	15.420	$\pm$	0.030	&	15.528	$\pm$	0.111	&	1										&	I,		P	\\
0132+254	&	19960 	&	7.45 	&	0.41	&	0.1	&	63	&	16.446	$\pm$	0.128	&		\nodata		&		\nodata		&	1,								9		&	  			\\
0816+297	&	16660 	&	7.84 	&	0.53	&	1.2	&	129	&	16.141	$\pm$	0.105	&	15.921	$\pm$	0.184	&		\nodata		&	1,									10	&	  			\\
0819+364	&	18740 	&	8.03 	&	0.64	&	2.3	&	123	&	16.026	$\pm$	0.088	&		\nodata		&	15.781	$\pm$	0.228	&	1,									10	&	  			\\
0821+633	&	16770 	&	7.82 	&	0.52	&	1.2	&	123	&	16.221	$\pm$	0.009	&	16.238	$\pm$	0.018	&	16.342	$\pm$	0.045	&	1										&	  		P	\\
0826+455	&	10370 	&	7.86 	&	0.52	&	1.2	&	513	&	15.001	$\pm$	0.006	&	14.947	$\pm$	0.009	&	14.927	$\pm$	0.018	&	1										&	I,	S,	P	\\
0852+659	&	19070 	&	8.13 	&	0.70	&	2.8	&	141	&	15.876	$\pm$	0.079	&	15.826	$\pm$	0.184	&		\nodata		&	1,							8			&	  			\\
0854+405	&	22250 	&	7.91 	&	0.58	&	1.7	&	46	&	15.419	$\pm$	0.009	&	15.496	$\pm$	0.014	&	15.553	$\pm$	0.029	&	1										&	I,		P	\\
0908+171	&	17340 	&	7.92 	&	0.57	&	1.6	&	129	&	16.403	$\pm$	0.013	&	16.458	$\pm$	0.024	&	16.573	$\pm$	0.064	&	1										&	  		P	\\
0915+526	&	15560 	&	7.96 	&	0.60	&	1.9	&	200	&	15.809	$\pm$	0.009	&	15.894	$\pm$	0.018	&	15.977	$\pm$	0.039	&	1										&	  		P	\\
0933+729	&	17380 	&	8.00 	&	0.62	&	2.1	&	148	&	16.074	$\pm$	0.009	&	16.067	$\pm$	0.018	&	16.108	$\pm$	0.045	&	1										&	  		P	\\
0938+286	&	14490 	&	7.82 	&	0.52	&	1.2	&	200	&	15.742	$\pm$	0.016	&	15.750	$\pm$	0.027	&	15.924	$\pm$	0.060	&	1										&	  		P	\\
0938+550	&	18530 	&	8.10 	&	0.68	&	2.6	&	148	&	15.153	$\pm$	0.025	&	15.306	$\pm$	0.033	&	15.396	$\pm$	0.053	&	1										&	I,		P	\\
0943+441	&	12820 	&	7.55 	&	0.41	&	0.1	&	339	&	13.643	$\pm$	0.028	&	13.707	$\pm$	0.041	&	13.722	$\pm$	0.046	&	1,								9		&	  			\\
0947+326	&	22060 	&	8.31 	&	0.82	&	3.9	&	117	&	16.022	$\pm$	0.015	&	16.044	$\pm$	0.023	&	15.841	$\pm$	0.069	&	1										&	I,		P	\\
0950+078	&	14770 	&	7.95 	&	0.59	&	1.8	&	229	&	16.148	$\pm$	0.011	&	16.280	$\pm$	0.021	&	16.280	$\pm$	0.049	&	1										&	  		P	\\
0954+697	&	21420 	&	7.91 	&	0.58	&	1.7	&	54	&	16.427	$\pm$	0.016	&	16.612	$\pm$	0.042	&	16.548	$\pm$	0.078	&	1										&	  		P	\\
0956+021	&	15670 	&	7.80 	&	0.51	&	1.1	&	151	&	16.007	$\pm$	0.012	&	16.073	$\pm$	0.022	&	16.130	$\pm$	0.049	&	1										&	  	S,	P	\\
1000$-$002	&	19470 	&	7.99 	&	0.62	&	2.1	&	98	&	16.474	$\pm$	0.019	&	16.558	$\pm$	0.054	&	16.614	$\pm$	0.094	&	1										&	  		P	\\
1003$-$023	&	20340 	&	7.95 	&	0.60	&	1.9	&	76	&	15.677	$\pm$	0.014	&	15.733	$\pm$	0.030	&	15.845	$\pm$	0.057	&	1										&	  		P	\\
1005+642	&	19660 	&	7.93 	&	0.58	&	1.7	&	81	&	14.232	$\pm$	0.029	&	14.330	$\pm$	0.050	&	14.356	$\pm$	0.052	&	1										&	  			\\
1015+161	&	19540	&	8.04 	&	0.65	&	2.3	&	110	&	15.950	$\pm$	0.010	&	15.944	$\pm$	0.018	&	15.906	$\pm$	0.036	&	1,			4							&	  		P	\\
1017+125	&	21670 	&	7.94 	&	0.60	&	1.9	&	56	&	16.236	$\pm$	0.096	&	15.856	$\pm$	0.177	&	15.875	$\pm$	0.253	&	1,									10	&	  			\\
1019+129	&	18020 	&	8.00 	&	0.62	&	2.1	&	132	&	16.101	$\pm$	0.017	&	16.048	$\pm$	0.031	&	15.981	$\pm$	0.058	&	1										&	I,		P	\\
1022+050	&	11680 	&	7.64 	&	0.44	&	0.4	&	490	&	14.193	$\pm$	0.035	&	14.237	$\pm$	0.033	&	14.185	$\pm$	0.065	&	1										&	  			\\
1026+024	&	12570 	&	7.98 	&	0.60	&	1.9	&	363	&	14.433	$\pm$	0.011	&	14.447	$\pm$	0.017	&	14.362	$\pm$	0.031	&	1										&	I,		P	\\
1031+063	&	20750 	&	7.87 	&	0.56	&	1.5	&	58	&	16.726	$\pm$	0.025	&	16.912	$\pm$	0.061	&	16.647	$\pm$	0.093	&	1										&	I,		P	\\
1031+234	&	15000 	&	8.50 	&	0.93	&	4.9	&	550	&	16.044	$\pm$	0.009	&	16.108	$\pm$	0.016	&	16.123	$\pm$	0.049	&	1										&	  		P	\\
1034+492	&	20650 	&	8.17 	&	0.73	&	3.1	&	11	&	15.778	$\pm$	0.063	&	16.127	$\pm$	0.212	&	16.061	$\pm$	0.287	&	1,							8			&	  			\\
1036+086	&	22230	&	7.49	&	0.42	&	0.2	&	46	&	16.654	$\pm$	0.015	&	16.810	$\pm$	0.034	&	16.471	$\pm$	0.066	&	1										&	I,	S,	P	\\
1046+282	&	12610 	&	7.97 	&	0.59	&	1.8	&	363	&	15.479	$\pm$	0.016	&	15.438	$\pm$	0.028	&	15.349	$\pm$	0.057	&	1										&	I,		P	\\
1102+749	&	19710 	&	8.36 	&	0.85	&	4.2	&	182	&	15.530	$\pm$	0.010	&	15.574	$\pm$	0.023	&	15.746	$\pm$	0.044	&	1										&	  		P	\\
1108+476	&	12400 	&	8.31 	&	0.80	&	3.7	&	603	&	15.657	$\pm$	0.054	&	15.350	$\pm$	0.092	&	15.618	$\pm$	0.222	&	1,							8			&	  			\\
1115+166	&	22090 	&	8.12 	&	0.70	&	2.8	&	81	&	15.466	$\pm$	0.055	&	15.657	$\pm$	0.159	&	15.463	$\pm$	0.208	&	1										&	  			\\
1116+026	&	12290 	&	8.05 	&	0.63	&	2.2	&	427	&	14.745	$\pm$	0.014	&	14.668	$\pm$	0.021	&	14.718	$\pm$	0.032	&	1,			4							&	  		P	\\
1122+546	&	14380 	&	7.83 	&	0.52	&	1.2	&	209	&	15.697	$\pm$	0.015	&	15.704	$\pm$	0.034	&	15.590	$\pm$	0.053	&	1										&	I,	S,	P	\\
1129+072	&	13360 	&	7.91 	&	0.56	&	1.5	&	288	&	15.298	$\pm$	0.012	&	15.158	$\pm$	0.018	&	15.266	$\pm$	0.033	&	1										&	  		P	\\
1129+156	&	16890 	&	8.19 	&	0.73	&	3.1	&	224	&	14.443	$\pm$	0.034	&	14.493	$\pm$	0.052	&	14.565	$\pm$	0.094	&	1,							8			&	  			\\
1134+301	&	21280 	&	8.55 	&	0.96	&	5.2	&	200	&	12.993	$\pm$	0.024	&	13.105	$\pm$	0.031	&	13.183	$\pm$	0.028	&	  		3								&	  			\\
1147+256	&	10200 	&	8.14 	&	0.69	&	2.7	&	794	&	15.548	$\pm$	0.012	&	15.555	$\pm$	0.020	&	15.668	$\pm$	0.047	&	1										&	  		P	\\
1149+058	&	11070 	&	8.15 	&	0.69	&	2.7	&	646	&	14.985	$\pm$	0.015	&	14.962	$\pm$	0.021	&	14.861	$\pm$	0.035	&	1										&	I,	S,	P	\\
1149+411	&	14070 	&	7.84 	&	0.53	&	1.2	&	224	&	16.365	$\pm$	0.017	&	16.431	$\pm$	0.037	&	16.398	$\pm$	0.064	&	1										&	  		P	\\
1158+433	&	14050 	&	7.85 	&	0.53	&	1.2	&	224	&	16.170	$\pm$	0.011	&	16.197	$\pm$	0.020	&	16.169	$\pm$	0.043	&	1										&	  		P	\\
1159$-$098	&	9540 	&	8.81 	&	1.10	&	6.5	&	2692	&	15.533	$\pm$	0.018	&	15.491	$\pm$	0.032	&	15.419	$\pm$	0.090	&	1										&	I,		P	\\
1201$-$001	&	19770 	&	8.26 	&	0.78	&	3.5	&	155	&	15.680	$\pm$	0.057	&	15.846	$\pm$	0.098	&	15.706	$\pm$	0.231	&	1,							8			&	  			\\
1216+036	&	13800 	&	7.85 	&	0.53	&	1.2	&	245	&	16.284	$\pm$	0.101	&	16.198	$\pm$	0.187	&		\nodata		&	1,									10	&	  			\\
1229$-$013	&	19430 	&	7.47 	&	0.41	&	0.1	&	71	&	14.925	$\pm$	0.032	&	15.044	$\pm$	0.070	&	15.349	$\pm$	0.229	&	1,								9		&	  			\\
1230+418	&	19520 	&	7.98 	&	0.61	&	2.0	&	95	&	16.161	$\pm$	0.014	&	16.180	$\pm$	0.027	&	16.337	$\pm$	0.055	&	1										&	  		P	\\
1232+479	&	14370 	&	7.82 	&	0.52	&	1.2	&	204	&	14.937	$\pm$	0.011	&	14.760	$\pm$	0.014	&	14.725	$\pm$	0.024	&	1										&	I,		P	\\
1237$-$029	&	10240 	&	8.58 	&	0.98	&	5.4	&	1660	&	15.828	$\pm$	0.011	&	15.761	$\pm$	0.018	&	15.955	$\pm$	0.040	&	1										&	  		P	\\
1244+149	&	10680 	&	8.06 	&	0.64	&	2.3	&	631	&	15.740	$\pm$	0.015	&	15.714	$\pm$	0.029	&	15.729	$\pm$	0.058	&	1										&	  		P	\\
1257+032	&	17290 	&	7.79 	&	0.51	&	1.1	&	105	&	16.090	$\pm$	0.092	&	15.956	$\pm$	0.159	&		\nodata		&	1,									10	&	  			\\
1258+593	&	14480	&	7.87	&	0.54	&	1.3	&	214	&	15.645	$\pm$	0.012	&	15.607	$\pm$	0.021	&	15.621	$\pm$	0.051	&	1										&	  		P	\\
1307+354	&	11180 	&	8.15 	&	0.70	&	2.8	&	631	&	15.344	$\pm$	0.051	&	15.398	$\pm$	0.096	&		\nodata		&	1,							8			&	  			\\
1310+583	&	10560 	&	8.32 	&	0.80	&	3.7	&	912	&	14.016	$\pm$	0.028	&	14.004	$\pm$	0.045	&	14.081	$\pm$	0.073	&	1,							8			&	  			\\
1319+466	&	13880 	&	8.19 	&	0.73	&	3.1	&	398	&	14.867	$\pm$	0.035	&	14.859	$\pm$	0.075	&	14.767	$\pm$	0.083	&	1,							8			&	  			\\
1325+168	&	17970 	&	8.18 	&	0.73	&	3.1	&	186	&	16.426	$\pm$	0.016	&	16.640	$\pm$	0.045	&	16.499	$\pm$	0.063	&	1										&	I,	S,	P	\\
1328+344	&	16300 	&	7.90 	&	0.56	&	1.5	&	155	&	15.495	$\pm$	0.008	&	15.528	$\pm$	0.015	&	15.654	$\pm$	0.030	&	1										&	  		P	\\
1330+473	&	22460 	&	7.95 	&	0.60	&	1.9	&	48	&	15.772	$\pm$	0.011	&	15.710	$\pm$	0.017	&	15.912	$\pm$	0.042	&	1										&	  		P	\\
1335+369	&	20510 	&	7.78 	&	0.51	&	1.1	&	51	&	15.122	$\pm$	0.010	&	15.152	$\pm$	0.016	&	15.188	$\pm$	0.029	&	1										&	  		P	\\
1337+705	&	20460 	&	7.90 	&	0.57	&	1.6	&	66	&	13.248	$\pm$	0.024	&	13.357	$\pm$	0.027	&	13.451	$\pm$	0.035	&	1,		3,								&	  			\\
1344+573	&	13390 	&	7.94 	&	0.58	&	1.7	&	295	&	13.704	$\pm$	0.022	&	13.821	$\pm$	0.035	&	13.757	$\pm$	0.041	&	1										&	  			\\
1349+144	&	16620 	&	7.68 	&	0.46	&	0.6	&	145	&	15.550	$\pm$	0.009	&	15.629	$\pm$	0.017	&	15.664	$\pm$	0.036	&	1										&	  		P	\\
1349+552	&	11800 	&	7.87 	&	0.54	&	1.3	&	380	&	15.830	$\pm$	0.015	&	15.967	$\pm$	0.028	&	15.800	$\pm$	0.060	&	1										&	I,		P	\\
1350+657	&	11880 	&	7.91 	&	0.55	&	1.4	&	389	&	15.688	$\pm$	0.068	&	15.419	$\pm$	0.136	&	15.304	$\pm$	0.175	&	1										&	  			\\
1350$-$090	&	9520 	&	8.36 	&	0.83	&	4.0	&	1318	&	14.191	$\pm$	0.007	&	14.082	$\pm$	0.009	&	14.140	$\pm$	0.013	&	1		   	   	   			   	   	   	&	  		P	\\
1407+425	&	10010 	&	8.21 	&	0.73	&	3.1	&	912	&	14.998	$\pm$	0.044	&	14.870	$\pm$	0.064	&	14.892	$\pm$	0.095	&	1,				5						&	  			\\
1408+324	&	18150	&	7.95	&	0.59	&	1.8	&	117	&	14.396	$\pm$	0.031	&	14.484	$\pm$	0.054	&	14.431	$\pm$	0.073	&	  		3								&	  			\\
1410+425	&	15350 	&	7.87 	&	0.54	&	1.3	&	178	&	16.343	$\pm$	0.016	&	16.337	$\pm$	0.025	&	16.467	$\pm$	0.069	&	1										&	  		P	\\
1431+154	&	13550 	&	7.95 	&	0.58	&	1.7	&	302	&	16.130	$\pm$	0.014	&	15.985	$\pm$	0.022	&	15.895	$\pm$	0.053	&	1										&	I,		P	\\
1448+078	&	14170 	&	7.75 	&	0.48	&	0.8	&	195	&	15.854	$\pm$	0.016	&	15.755	$\pm$	0.027	&	15.788	$\pm$	0.072	&	1										&	I,		P	\\
1449+168	&	21600 	&	7.88 	&	0.56	&	1.5	&	49	&	15.973	$\pm$	0.010	&	16.077	$\pm$	0.018	&	15.861	$\pm$	0.044	&	1										&	I,		P	\\
1457$-$086	&	21450 	&	7.97 	&	0.62	&	2.1	&	63	&	16.041	$\pm$	0.098	&	16.212	$\pm$	0.233	&	15.614	$\pm$	0.228	&	1,						7				&	  			\\
1501+032	&	13770 	&	7.88 	&	0.54	&	1.3	&	257	&	15.856	$\pm$	0.018	&	15.847	$\pm$	0.033	&	15.900	$\pm$	0.063	&	1										&			P	\\
1502+351	&	18120 	&	8.13 	&	0.70	&	2.8	&	166	&	16.239	$\pm$	0.024	&	16.386	$\pm$	0.038	&	16.305	$\pm$	0.072	&	1										&	I,		P	\\
1507+021	&	19580 	&	7.87 	&	0.56	&	1.5	&	76	&	16.601	$\pm$	0.115	&	16.451	$\pm$	0.209	&		\nodata		&	1,									10	&				\\
1507+220	&	19340 	&	7.91 	&	0.57	&	1.6	&	85	&	15.445	$\pm$	0.015	&	15.525	$\pm$	0.021	&	15.698	$\pm$	0.055	&	1										&			P	\\
1508+549	&	16970 	&	7.86 	&	0.54	&	1.3	&	126	&	16.125	$\pm$	0.009	&	16.136	$\pm$	0.019	&	16.278	$\pm$	0.051	&	1										&			P	\\
1508+637	&	10450 	&	8.12 	&	0.68	&	2.6	&	724	&	14.689	$\pm$	0.015	&	14.660	$\pm$	0.023	&	14.823	$\pm$	0.046	&	1										&			P	\\
1509+323	&	13970	&	7.98	&	0.60	&	1.9	&	282	&	14.436	$\pm$	0.030	&	14.574	$\pm$	0.055	&	14.572	$\pm$	0.089	&			3								&				\\
1515+669	&	10320 	&	8.40 	&	0.86	&	4.3	&	1122	&	15.295	$\pm$	0.053	&	15.240	$\pm$	0.112	&	15.180	$\pm$	0.198	&	1,							8			&				\\
1519+384	&	19620 	&	7.98 	&	0.61	&	2.0	&	93	&	16.207	$\pm$	0.015	&	16.359	$\pm$	0.032	&	16.176	$\pm$	0.080	&	1										&	I,		P	\\
1527+091	&	21520 	&	8.02 	&	0.64	&	2.3	&	71	&	14.814	$\pm$	0.012	&	14.922	$\pm$	0.018	&	15.003	$\pm$	0.037	&	1										&			P	\\
1531$-$023	&	18620	&	8.41	&	0.88	&	4.5	&	234	&	14.395	$\pm$	0.040	&	14.484	$\pm$	0.053	&	14.618	$\pm$	0.101	&			3								&				\\
1533$-$057	&	20000 	&	8.50 	&	0.94	&	5.0	&	245	&	15.876	$\pm$	0.013	&	15.951	$\pm$	0.031	&	16.082	$\pm$	0.059	&	1										&			P	\\
1537+652	&	9740 	&	8.15 	&	0.69	&	2.7	&	912	&	14.438	$\pm$	0.006	&	14.426	$\pm$	0.009	&	14.372	$\pm$	0.014	&	1										&	I		P	\\
1541+651	&	11600 	&	8.10 	&	0.67	&	2.5	&	537	&	15.625	$\pm$	0.013	&	15.590	$\pm$	0.025	&	15.450	$\pm$	0.048	&	1										&	I,	S,	P	\\
1548+149	&	20520 	&	7.89 	&	0.57	&	1.6	&	65	&	15.656	$\pm$	0.013	&	15.844	$\pm$	0.029	&	15.821	$\pm$	0.061	&	1										&	I		P	\\
1550+183	&	14260 	&	8.25 	&	0.77	&	3.4	&	389	&	15.170	$\pm$	0.050	&	15.311	$\pm$	0.107	&	15.278	$\pm$	0.131	&	1,							8			&				\\
1601+581	&	14670 	&	7.84 	&	0.53	&	1.2	&	200	&	14.116	$\pm$	0.010	&	14.151	$\pm$	0.013	&	14.209	$\pm$	0.021	&	1										&			P	\\
1605+684	&	21090 	&	7.97 	&	0.61	&	2.0	&	68	&	16.404	$\pm$	0.010	&	16.474	$\pm$	0.026	&	16.276	$\pm$	0.054	&	1										&		S,	P	\\
1610+167	&	14390 	&	7.84 	&	0.52	&	1.2	&	209	&	16.047	$\pm$	0.011	&	16.033	$\pm$	0.024	&	16.027	$\pm$	0.065	&	1										&	I,		P	\\
1614+137	&	22430 	&	7.33 	&	0.40	&	0.1	&	30	&	15.704	$\pm$	0.059	&	15.896	$\pm$	0.141	&		\nodata		&	1										&				\\
1620+513	&	20890 	&	7.92 	&	0.58	&	1.7	&	63	&	16.251	$\pm$	0.093	&	16.181	$\pm$	0.213	&		\nodata		&	1,									10	&				\\
1626+409	&	21370 	&	8.02 	&	0.64	&	2.3	&	72	&	16.647	$\pm$	0.027	&	16.933	$\pm$	0.068	&	16.851	$\pm$	0.146	&	1										&	I,		P	\\
1632+177	&	10100 	&	7.96 	&	0.58	&	1.7	&	617	&	13.049	$\pm$	0.021	&	12.989	$\pm$	0.027	&	13.076	$\pm$	0.029	&	1,				5						&				\\
1637+335	&	10150	&	8.17	&	0.71	&	2.9	&	832	&	14.551	$\pm$	0.031	&	14.467	$\pm$	0.045	&	14.424	$\pm$	0.081	&			3								&				\\
1641+388	&	15570 	&	7.95 	&	0.59	&	1.8	&	195	&	14.958	$\pm$	0.008	&	15.066	$\pm$	0.014	&	15.024	$\pm$	0.037	&	1										&	I,		P	\\
1647+376	&	21980 	&	7.89 	&	0.57	&	1.6	&	46	&	15.494	$\pm$	0.007	&	15.533	$\pm$	0.016	&	15.782	$\pm$	0.051	&	1										&	I,		P	\\
1654+637	&	15070 	&	7.63 	&	0.44	&	0.4	&	204	&	16.167	$\pm$	0.103	&	15.688	$\pm$	0.146	&	15.565	$\pm$	0.179	&	1,								9		&				\\
1659+303	&	13600	&	7.95	&	0.58	&	1.7	&	295	&	15.315	$\pm$	0.011	&	15.328	$\pm$	0.025	&	15.435	$\pm$	0.045	&	1										&	I		P	\\
1720+361	&	13670 	&	7.83 	&	0.52	&	1.2	&	245	&	15.476	$\pm$	0.023	&	15.592	$\pm$	0.046	&	15.136	$\pm$	0.075	&	1										&	I,	S,	P	\\
2226+061	&	15280 	&	7.62 	&	0.44	&	0.4	&	191	&	15.178	$\pm$	0.045	&	15.217	$\pm$	0.067	&	15.227	$\pm$	0.142	&	1,								9		&				\\
2246+223	&	10650	&	8.80	&	1.10	&	6.5	&	2188	&	14.341	$\pm$	0.029	&	14.317	$\pm$	0.047	&	14.360	$\pm$	0.090	&			3								&				\\
2303+243	&	11480 	&	8.09 	&	0.66	&	2.4	&	550	&	15.398	$\pm$	0.028	&	15.509	$\pm$	0.042	&	15.363	$\pm$	0.073	&	1										&		S,	P	\\
2306+125	&	20220 	&	8.05 	&	0.66	&	2.4	&	98	&	15.672	$\pm$	0.012	&	15.642	$\pm$	0.024	&	15.735	$\pm$	0.068	&	1										&	I,		P	\\
2306+131	&	13250 	&	7.92 	&	0.56	&	1.5	&	295	&	15.567	$\pm$	0.013	&	15.553	$\pm$	0.026	&	15.299	$\pm$	0.065	&	1										&	I,	S,	P	\\
2322+207	&	13060 	&	7.84 	&	0.52	&	1.2	&	282	&	15.745	$\pm$	0.013	&	15.708	$\pm$	0.026	&	15.354	$\pm$	0.066	&	1										&	I,		P	\\
2324+060	&	15750 	&	7.90 	&	0.56	&	1.5	&	174	&	15.744	$\pm$	0.028	&	15.879	$\pm$	0.063	&	16.051	$\pm$	0.249	&	1										&	I,		P	\\
2326+049	&	11820 	&	8.15 	&	0.70	&	2.8	&	550	&	13.132	$\pm$	0.029	&	13.075	$\pm$	0.022	&	12.689	$\pm$	0.029	&	1,	2,	3,		5,	6					&				\\
2328+108	&	21910 	&	7.84 	&	0.55	&	1.4	&	42	&	15.900	$\pm$	0.013	&	15.911	$\pm$	0.023	&	16.046	$\pm$	0.109	&	1										&	I,		P	\\
2329+267	&	11730 	&	8.98 	&	1.18	&	7.2	&	2042	&	15.126	$\pm$	0.011	&	15.093	$\pm$	0.020	&	14.908	$\pm$	0.088	&	1										&	I,		P	\\
2336+063	&	16520 	&	8.03 	&	0.63	&	2.2	&	186	&	15.903	$\pm$	0.017	&	15.959	$\pm$	0.032	&	16.140	$\pm$	0.135	&	1										&	I,		P	
\enddata								
\tablecomments{Data sources: (I) IRTF; (S) \emph{Spitzer}; (P) PAIRITEL}										
\tablerefs{
(1) this work;
(2) \citealt{reach}; 
(3) \citealt{mullally}; 
(4) \citealt{jura}; 
(5) \citealt{farihi}; 
(6) \citealt{reach09}; 
(7) \citealt{farihi09}; 
(8) \citealt{gould}; 
(9) \citealt{kilic}; 
(10) \citealt{xu}}\label{table1}
\end{deluxetable}
%
%
\begin{deluxetable}{ccc}
\tabletypesize{\footnotesize}
\tablewidth{0pt}
\tablecaption{\emph{Spitzer} IRAC Photometry}
\tablehead{									
	&	3.6 $\mu$m			&	4.5 $\mu$m			\\
PG	&		($\mu$Jy)		&	($\mu$Jy)			\\
}									
\startdata									
0826+455	&	268.6	$\pm$	8.7	&	172.0	$\pm$	5.9	\\
0956+021	&	88.6	$\pm$	3.5	&	53.7	$\pm$	2.8	\\
1036+086	&	42.5	$\pm$	1.5	&	26.6	$\pm$	1.3	\\
1122+546	&	118.2	$\pm$	4.3	&	77.6	$\pm$	3.1	\\
1149+058	&	248.0	$\pm$	8.1	&	163.4	$\pm$	5.7	\\
1325+168	&	64.7	$\pm$	2.2	&	42.3	$\pm$	1.7	\\
1541+651	&	435.4	$\pm$	13.5	&	483.9	$\pm$	15	\\
1605+684	&	60.9	$\pm$	2.7	&	41.2	$\pm$	2.2	\\
1720+361	&	155.9	$\pm$	5.4	&	97.6	$\pm$	3.7	\\
2303+243	&	170.8	$\pm$	5.9	&	110.1	$\pm$	4	\\
2306+131	&	146.2	$\pm$	5.2	&	92.0	$\pm$	3.5	
\enddata									
\tablecomments{\emph{Spitzer} Cycle 7 photometry fluxes in IRAC Channels 1 (3.6 $\mu$m) and 2 (4.5 $\mu$m).}
\label{table2}																		
\end{deluxetable}
%
%
\end{document}